\documentclass[twocolumn,tighten]{aastex63}

\usepackage{amsmath}
\usepackage{txfonts}
\usepackage{graphicx}
\usepackage{comment}
\usepackage{hyperref}
\usepackage[utf8]{inputenc}
\usepackage{multirow}
\usepackage{dcolumn}

\usepackage{xcolor}

\shorttitle{Is the  neutrino event IC200530A  associated with a hydrogen rich superluminous supernova?}
\shortauthors{Pitik et al.}

\begin{document}

\title{Is the high-energy neutrino event IceCube-200530A  associated with a hydrogen rich superluminous supernova?}

\author[0000-0002-9109-2451]{Tetyana Pitik}
\affiliation{Niels Bohr International Academy, Niels Bohr Institute, University of Copenhagen, Blegdamsvej 17, 2100, Copenhagen, Denmark}
\affiliation{DARK, Niels Bohr Institute, University of Copenhagen, Jagtvej 128, 2200, Copenhagen, Denmark}
\author[0000-0001-7449-104X]{Irene Tamborra}
\affiliation{Niels Bohr International Academy, Niels Bohr Institute, University of Copenhagen, Blegdamsvej 17, 2100, Copenhagen, Denmark}
\affiliation{DARK, Niels Bohr Institute, University of Copenhagen, Jagtvej 128, 2200, Copenhagen, Denmark}
\author[0000-0002-4269-7999]{Charlotte R. Angus} 
\affiliation{DARK, Niels Bohr Institute, University of Copenhagen, Jagtvej 128, 2200, Copenhagen, Denmark}
\author[0000-0002-4449-9152]{Katie Auchettl}
\affiliation{School of Physics, The University of Melbourne, Parkville, VIC 3010, Australia}
\affiliation{ARC Centre of Excellence for All Sky Astrophysics in 3 Dimensions (ASTRO 3D)}
\affiliation{Department of Astronomy and Astrophysics, University of California, Santa Cruz, CA 95064, USA}

\begin{abstract}
The Zwicky Transient Facility (ZTF) follow-up campaign of alerts released by the IceCube Neutrino Observatory  has led to the likely identification of the transient AT2019fdr as the source of  the neutrino event  IC200530A. AT2019fdr was initially suggested to be a tidal disruption event in a Narrow-Line Seyfert 1 galaxy. However, the combination of its spectral properties,  color evolution, and feature-rich light curve suggests that AT2019fdr may be a Type IIn superluminous supernova.
In the latter scenario, IC200530A may have been produced via inelastic proton-proton collisions between the relativistic protons accelerated at the forward shock and the cold protons of the circumstellar medium. Here, we investigate this possibility and find that at most  $4.6\times 10^{-2}$
muon neutrino and antineutrino events are expected to be detected by the IceCube Neutrino Observatory within $394$~days of discovery in the case of excellent  discrimination of the atmospheric background. After correcting for the Eddington bias, which occurs when a single cosmic neutrino event is adopted to infer the neutrino emission at the source, we conclude that  IC200530A may originate from the hydrogen rich superluminous supernova AT2019fdr.
\end{abstract}

\keywords{type II supernovae --- neutrino astronomy --- particle astrophysics}

\section{Introduction}

In 2013, the IceCube Collaboration reported the detection of a flux of high-energy neutrinos of astrophysical origin, marking  the beginning of the high-energy neutrino astronomy era. Despite the growing number of  high-energy neutrino events  detected by the IceCube Neutrino Observatory,  the sources   of the cosmic neutrino flux remain to be unveiled~\citep{IceCube:2018fhm,IceCube:2021xar,IceCube:2020wum,IceCube:2020acn}. 

High energy neutrino events have been reported to be in likely coincidence with blazars~\citep{IceCube:2018dnn,Giommi:2020viy,Franckowiak:2020qrq,Fermi-LAT:2019hte,Krauss:2018tpa,Kadler:2016ygj}. However, association studies of blazars hint towards no excess from the broader population~\citep{IceCube:2016qvd}. Various other source classes have been proposed as factories of the  observed cosmic neutrino flux~\citep{Meszaros:2017fcs,Ahlers:2018fkn,Vitagliano:2019yzm}, such as gamma-ray bursts, cluster of galaxies, star-forming galaxies, and tidal distruption events~\citep{Meszaros:2015krr,Pitik:2021xhb,Murase:2015ndr,Waxman:2015ues,Tamborra:2014xia,Zandanel:2014pva,Wang:2015mmh,Dai:2016gtz,Senno:2016bso,Lunardini:2016xwi}. Nevertheless, the neutrino emission from each of  these source classes cannot  fully account for  the observed neutrino flux. 

The growing number of cosmic neutrino alerts has triggered follow-up searches for coincident detection of electromagnetic radiation, see e.g.~\cite{IceCube:2020mzw,Fermi-LAT:2019hte,Acciari:2021YA}. On October 1st 2019, the IceCube Collaboration reported the detection of a muon track neutrino  of likely astrophysical origin, IC191001A. This event has been suggested to be the neutrino counterpart of the  tidal distruption event (TDE) candidate AT2019dsg which was discovered by the Zwicky Transient Facility (ZTF) -- see e.g.~\cite{Stein:2020xhk,2019PASP..131a8002B}. Various theoretical models have been discussed to interpret this likely association~\citep{Winter:2020ptf,Liu:2020isi,Murase:2020lnu}, however the jetted version of these models is being challenged by the most recent work on the radio properties of AT2019dsg~\citep{2021arXiv210306299C,2021arXiv210615799M,2021MNRAS.507.4196M,2021arXiv210902648M}.

More recently, the follow-up campaign of IceCube neutrino alerts carried out by the ZTF Collaboration has led to another transient association. On May 31st 2020, \citet{2020GCN.27865....1I,2020GCN.27872....1R} detected another muon track candidate (IC200530A), which was suggested to be associated with the optical transient AT2019fdr/ZTF19aatubsj\footnote{\url{https://www.wis-tns.org/object/2019fdr}} located at redshift $z=0.2666$.
The IC200530A event was detected $\sim394$~days after the discovery of the transient (hereafter identified with the onset of the shock breakout) and about $300$~days after the peak of the electromagnetic emission. This neutrino event has a reconstructed neutrino energy of $E_{\nu}\simeq 80 $~TeV and  a signalness larger than $50 \%$~\citep{2020GCN.27865....1I,2020GCN.27872....1R,Stein:2021FH}.

The intriguing  coincidence of two IceCube neutrino events with two ZTF transient sources has  triggered  searches by the ANTARES Collaboration~\citep{ANTARES:2021jmp} and led to stringent upper limits on the neutrino emission from both sources. In addition, the Baikal-GVD Collaboration is currently investigating  clusters of neutrino events detected along the same angular directions of both ZTF sources~\citep{Suvorova:2021ou}.

AT2019fdr is located close to the nucleus of its host galaxy and shows strong narrow hydrogen emission lines within its spectra. This led to the initial classification of AT2019fdr as either a flaring active galactic nucleus (AGN) in a Narrow-Line Seyfert 1 galaxy~\citep{2020arXiv201008554F}, or a tidal disruption event~\citep{Chornock2019}. This has resulted in interpretations of IC200530A being associated with an accreting black hole transient event~\citep{Stein:2021FH}. However, \cite{Yan:2019TNSAN} proposed that AT2019fdr is a hydrogen-rich superluminous supernova (otherwise named superluminous supernova of Type IIn, SLSN IIn). Hydrogen rich SLSNe exhibit strong narrow Balmer emission lines within their spectra, but are more luminous than standard type IIn supernovae (SNe IIn), achieving luminosities typically with $M \lesssim -20$ at peak brightness~\citep{Gal-Yam:2012ukv,Smith:2014txa,Gal-Yam2019}. The narrow emission lines within SNe IIn are indicative of interaction between the SN ejecta with a dense shell of surrounding circumstellar material (CSM) in which kinetic energy is efficiently converted into thermal energy. The high luminosity of SLSNe IIn is thought to be the result of either an highly energetic explosion [with typical energies $E_{\rm{ej}}\simeq \mathcal{O}(10^{51}$--$10^{52})$~ergs], interaction with an unusually massive CSM~\citep{Moriya:2018sig}, or some combination of the two scenarios.

Proton acceleration, even beyond PeV energies, could take place in the proximity of the SLSN shock expanding in the dense CSM. The interaction of these protons with those of the shocked CSM may lead to copious neutrino emission~\citep{Murase:2010cu,Katz:2011zx,2014MNRAS.440.2528M,Cardillo:2015zda,Zirakashvili:2015mua,Petropoulou:2016zar,Petropoulou:2017ymv,Murase:2017pfe}. In this work, we investigate the possibility that IC200530A originates from AT2019fdr, under the framework that this transient is a SLSN IIn. 

This paper is organized as follows. After introducing the main features characterizing AT2019fdr in Sec.~\ref{sec:AT2019fdr}, we outline the setup adopted to predict the neutrino signal in Sec.~\ref{sec:setup}. Our findings are presented in Sec.~\ref{sec:results} together with a discussion on the dependence of the neutrino signal on the uncertain parameters characteristic of AT2019fdr. A discussion on our findings and caveats is reported in Sec.~\ref{sec:discussion}, followed by our conclusions in Sec.~\ref{sec:conclusions}. In addition, Appendix~\ref{Appendix: parameter space} discusses how  the AT2019fdr parameter space  is  constrained by the observational constraints on AT2019fdr that we apply from neutrino and electromagnetic data. We investigate the temporal evolution of the maximum proton energy as a function of the model parameters considered for AT2019fdr in Appendix~\ref{Appendix: maximum_proton_energy}.

\section{AT2019fdr: A type IIn superluminous supernova}
\label{sec:AT2019fdr}

AT2019fdr exhibits many properties compatible with those of other documented SLSNe IIn from the literature. Spectroscopically, the event shows intermediate-width ($\sim$1000 km s$^{-1}$) Balmer emission lines combined with narrow galaxy emission lines from the host, superimposed upon a blue continuum\footnote{The classification spectra are publicly available on  \url{https://www.wiserep.org/object/12537}}. The intermediate width Balmer emission features are characteristic of interacting core-collapse SNe (SNe IIn and SLSNe IIn), see e.g.~\cite{2017hsn..book..195G,Moriya:2018sig}. Although these lines are also observed within nuclear transients (AGN flares and TDEs), the lack of intermediate components to the other host galaxy emission features (e.g.~O III) disfavors the interpretation of this transient as an AGN flare. It is unlikely that these features mark AT2019fdr as a TDE, as these events generally exhibit much broader emission profiles than seen in AT2019fdr (typically line widths $\sim 10^4$ km s$^{-1}$, \citealt{2017ApJ...842...29H, Charalampopoulos2021}).

\begin{figure}
	\centering
	\includegraphics[width=0.48\textwidth]{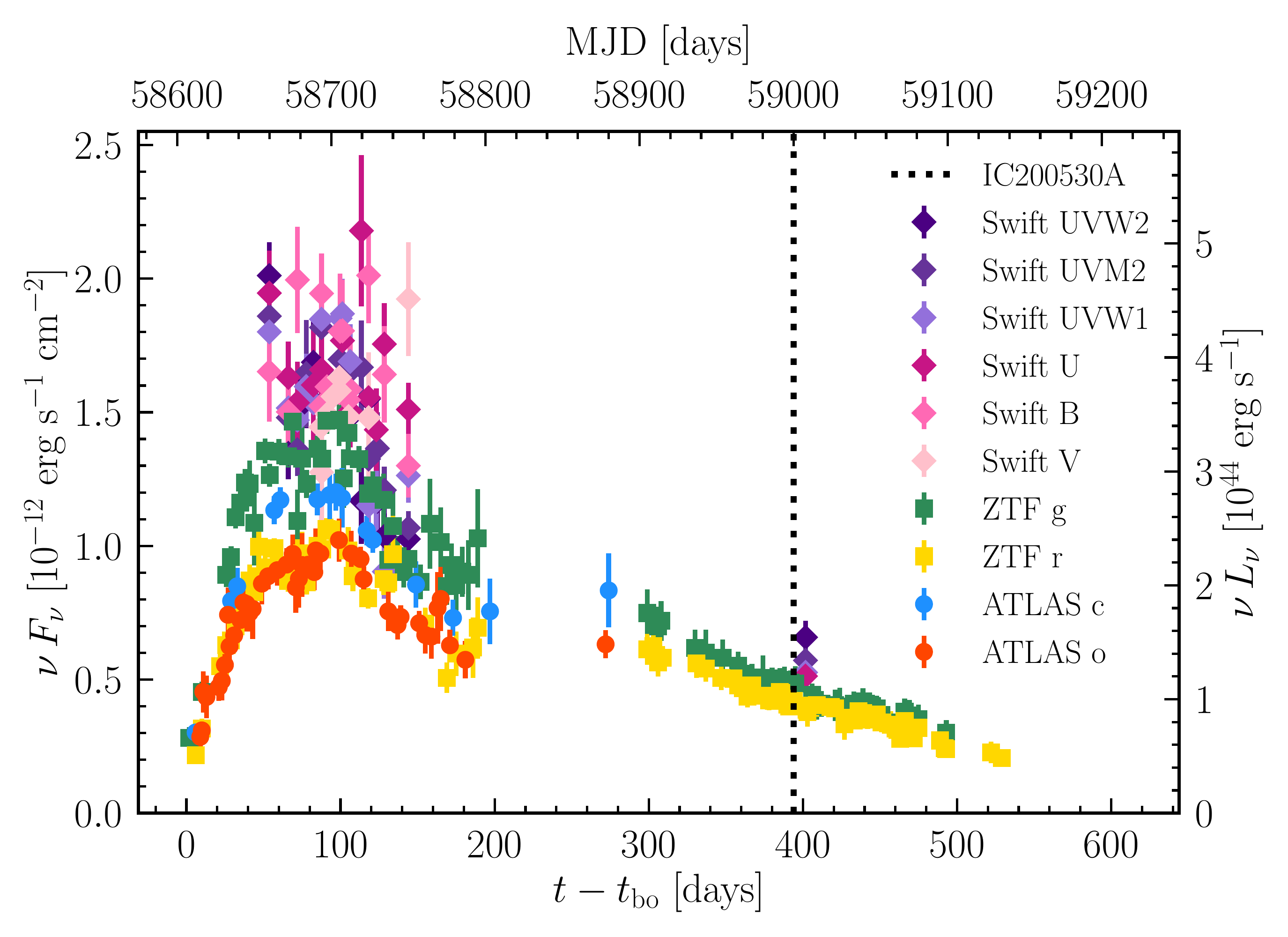}
	\caption{Ultraviolet-optical light curve of AT2019fdr. Public data taken from  ZTF \citep{Patterson2019}, ATLAS \citep{ATLAS2018,ATLAS2020} and \textit{Swift} \citep{gehrels04}. The detection epoch of IC200530A is marked as the black dashed vertical line and was observed $\sim 394$~days after the first optical detection of the SN in the observer frame. We display the time from estimated shock breakout ($t_{\rm{bo}}$), along the $x$-axis.
    }
	\label{Fig:LC_SN}
\end{figure}

The photometric behavior of AT2019fdr shows several features within the multi-band light curve, as displayed in Fig.~\ref{Fig:LC_SN}, consistent with interaction-powered SNe. Although the slow rise time ($\sim$80 days in the rest frame) and lengthy decline of the transient can be interpreted under each of the three potential paradigms suggested for its origin, the photometric evolution of AT2019fdr is not smooth. The light curve has a clear bump close to the peak (around 60 days from first light in the rest frame) alongside the beginning of an apparent re-brightening feature around $70$~days after the optical peak. Episodes of re-brightening have been observed within some SNe IIn \citep[e.g.][]{Stritzinger2012,Nyholm2017} and are attributed to changes in the CSM density and variable progenitor mass-loss rates. 

The late-time evolution of the transient ($> 160$~days from peak brightness) exhibits a slower decline than either Co$^{56}$-decay (from a standard Ni$^{56}$ powered light curve) or the $\alpha=-5/3$ power-law decline predicted by models of fallback accretion in TDEs \citep[e.g.][]{Rees1988}, but consistent with the range of typically slow declines in interacting SNe \citep{Nyholm2020}. AT2019fdr also exhibits a gradual reddening in color from peak to late times (transitioning from g - r $\sim -0.12$--$0.2$~mag), a property not predicted in nuclear transients, which shows an almost constant optical color in the majority of their light curves, but accordant with observations of normal SN IIn \citep{Taddia2013}. Finally, pre-explosion variability is also not observed within the ZTF and ATLAS imaging~\citep{Yan:2019TNSAN}, which disfavors an AGN origin.

Given the redshift of AT2019fdr, it is not possible to recover its complete rise in the ZTF photometry. However, non-detections in the ZTF g-band prior to first light place the breakout epoch $6$~days ($5$ rest frame days) before the first ZTF detection (see Fig.~\ref{Fig:LC_SN}). Fitting a low order polynomial to the rise of the ZTF curve suggests  that the start of the optical light curve coincides with these non-detections. Based on this, we assume the onset of the shock breakout at the first detection of MJD $= 58606 \pm 6$~days.

We also note that AT2019fdr was not the only source suggested to be associated with the neutrino event IC200530A. AT2020lam~\footnote{\url{https://wis-tns.weizmann.ac.il/object/2020lam}} and AT2020lls~\footnote{\url{https://wis-tns.weizmann.ac.il/object/2020lls}} were also suggested to be possibly associated, as they were found within a $90.0\%$ localization of the neutrino event~\citep{2020GCN.27872....1R}. AT2020lam was classified using the Nordic Optical Telescope as a Type II SN located at  $z=0.033$~\citep{2020GCN.27910....1R}. However, the spectrum and light curve showed no evidence of CSM interaction, necessary for neutrino producing, leading \cite{2020GCN.27910....1R} to suggest that it was not associated with the neutrino event IC200530A.  

AT2020lls was also classified using the Nordic Optical Telescope, but as a Type Ic SN located at  $z=0.04106$ that occurred $\sim 8$~days prior to the detection of IC200530A~\citep{2020GCN.27980....1R}. As this source did not show broad absorption features consistent with a subclass of Type Ic SN called Type Ic-BL, which are commonly associated with  off-axis gamma-ray bursts or  choked jets, \citet{2020GCN.27980....1R} suggested this was not associated with the neutrino event IC200530A.

\section{Model setup}
\label{sec:setup}

In this section, we introduce the method adopted to compute the neutrino spectral energy distribution from AT2019fdr and its temporal evolution, as well as the properties of AT2019fdr useful to this purpose. Details on the estimation of the neutrino flux and event rate expected at Earth follow.

\subsection{Spectral energy distributions of protons and neutrinos}
\label{sec:energy_distributions}

\begin{figure}
	\centering
	\includegraphics[width=0.45\textwidth]{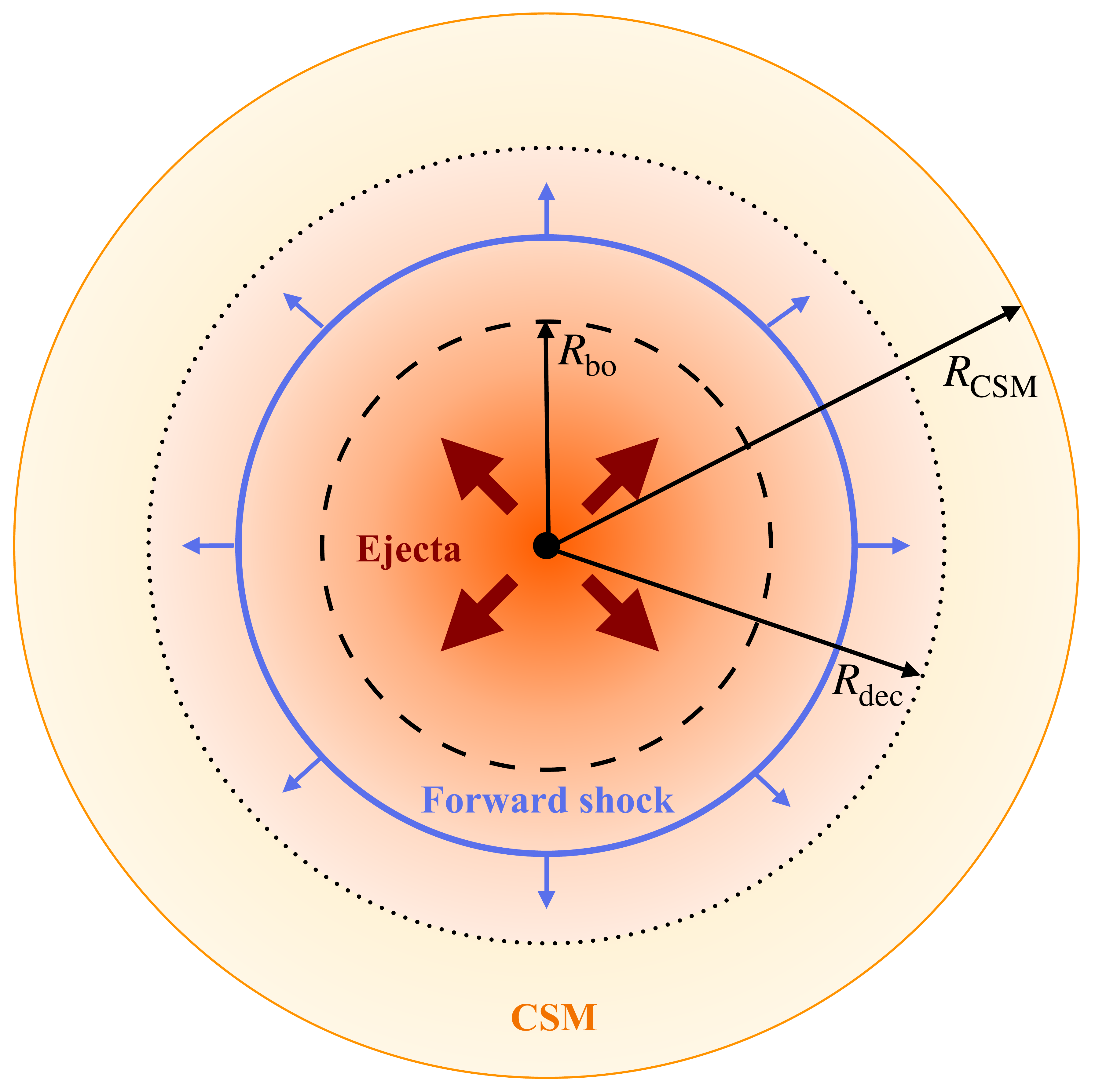}
	\caption{Schematic representation of AT2019fdr  after the explosion, assuming spherical symmetry. The central compact object (in black) is surrounded by the SN ejecta (orange region, with the bordeaux arrows indicating the  propagation of the ejected material) and a dense CSM envelope (yellow region) which extends up to its outer edge marked by $R_{\rm{CSM}}$. The color gradient describes the density
    gradient (from darker to lighter hues as the density decreases). The dashed
    black line  marks the position of the breakout radius ($R_{\rm{bo}}$). The indigo line represents the forward shock that propagates radially outwards. The  black dotted line marks the location of the deceleration radius of the ejecta ($R_{\rm{dec}}$). The latter is located at radii smaller  than $R_{\rm{CSM}}$ (as in this sketch) for a relatively large CSM mass compared to the ejecta mass or larger than   $R_{\rm{CSM}}$ for very massive ejecta and rarefied CSM; see Eq.~\ref{eq:Rdec}. For extremely large $M_{\rm{CSM}}/M_{\rm{ej}}$, it is possible that $R_{\rm{dec}}<R_{\rm{bo}}$.
    }
	\label{Fig:sketch_SN}
\end{figure}

We assume  a  spherical, steady and wind-like circumstellar medium (CSM) with solar composition ejected from the massive progenitor in the final stages of its evolution, as sketched in Fig.~\ref{Fig:sketch_SN}.  We define its  number density profile as
\begin{equation}
\label{eq:n_CSM}
n_{\rm{CSM}}(R)=\frac{\rho_{\rm{CSM}}(R)}{m}=\frac{\dot{M}}{4\pi v_{w} m R^{2}}\ ,
\end{equation}
where $ \dot{M} $ is the stellar mass loss rate, $ v_{w}$ the wind velocity, $ m=\mu m_{\rm{H}}, $ with $\mu= 1.3$ being the mean molecular weight for a neutral gas of solar abundance, and $R$ the distance to the stellar core.

The interaction of the stellar ejecta with the CSM leads to the formation of a forward shock (propagating in the CSM) and a reverse shock (propagating back into the stellar ejecta). Both the forward and reverse shocks could, in principle, contribute to the neutrino emission. Working under the assumption that the ejecta density profile decreases steeply~\citep{2003LNP...598..171C}, we neglect the contribution of the reverse shock since the forward shock is expected to predominantly contribute to the total energy dissipation rate and dominate the particle acceleration observed in SN remnants \citep[e.g.,][]{2007ApJ...661..879E, Patnaude:2008gq, 2010MNRAS.406.2633S, 2015ApJ...799..238S,2018ApJ...853...46S,Suzuki:2020qui}. Hence, we focus on the neutrino emission from the forward shock for the sake of simplicity. 

Following \cite{1982ApJ...258..790C,Moriya:2013hka}, we assume that  spherically symmetric SN ejecta of mass $M_{\rm{ej}}$ and kinetic energy $E_{k}$ expand  in the surrounding CSM homologously. The CSM extends up to an external radius $R_{\rm{CSM}}$ (see Fig.~\ref{Fig:N_numu_rate_band}). The outer ejecta density profile, which is relevant  for the interactions leading to neutrino production, scales  as $n_{\rm{ej}}\propto R^{-s}$, where we assume $s=10$. The shocked SN ejecta and CSM form a thin dense shell because of  efficient radiative cooling. Being the thickness of the thin shocked shell much smaller than its radius, one can describe its evolution through the radius $R_{\rm{sh}}(t)$. In the ejecta dominated phase, namely in the phase in which most part of the ejecta is still freely expanding (i.e., when the mass of the ejecta is larger than the swept-up CSM mass), the shock radius is given by~\citep{ Moriya:2013hka,Chevalier:2016hzo}:
\begin{equation}
\label{Eq: Rsh in free phase}
    R_{\rm{sh}}(t)=\bigg[\frac{2}{s(s -4)(s -3)}\frac{[10(s-5)E_{\rm{k}}]^{(s-3)/2}}{[3 (s-3) M_{\rm{ej}}]^{(s-5)/2}}\frac{v_{w}}{\dot{M}}\bigg]^{{1}/{(s -2)}} t^{{(s-3)}/{(s -2)}}\ ,
\end{equation}
with the corresponding shock velocity $v_{\rm{sh}}={\rm{d}} R_{\rm{sh}}/{\rm{d}}t$.

Because of the high CSM density, the forward shock is initially expanding in a radiation dominated region, and particle acceleration is not efficient~\citep{1976ApJS...32..233W,Levinson:2007rj,Katz:2011zx,Murase:2010cu}. Efficient particle acceleration takes place at radii larger than that of the shock breakout ($R_{\rm{bo}}$), where initially trapped photons are free to diffuse out to the photosphere; the shock breakout radius  is computed by solving the following equation:
\begin{equation}
{\tau_{T}(R_{\rm{bo}})=\int_{R_{\rm{bo}}}^{R_{\rm{CSM}}} \rho_{\rm{CSM}}(R) \kappa_{\rm{es}} dR = \frac{c}{v_{\rm{sh}}}}\ ,
\end{equation}
where $ \kappa_{\rm{es}}\sim 0.34\, \rm{cm}^{2} \rm{g}^{-1}$~\citep{Pan:2013nfa} is the electron scattering opacity at solar abundances, and $ c $ is the speed of light.
When the SN ejecta mass $M_{\rm{ej}}$ becomes comparable to the swept-up mass from the CSM, the ejecta enters the CSM-dominated phase. This transition happens at the deceleration radius
\begin{equation}
\label{eq:Rdec}
	R_{\rm{dec}} =\frac{M_{\rm{ej}}v_{w}}{\dot{M}}\ .
\end{equation}
Note that $R_{\rm{dec}}$ may be located at radii smaller than $R_{\rm{CSM}}$ as shown in Fig.~\ref{Fig:sketch_SN}, or larger than $R_{\rm{CSM}}$ according to the relative ratio between $M_{\rm{ej}}$ and $M_{\rm{CSM}}$ (i.e., if $M_{\rm{CSM}} > M_{\rm{ej}}$, then $R_{\rm{dec}}  < R_{\rm{CSM}}$  and viceversa). Furthermore, for  $M_{\rm{CSM}}$ extremely large with respect to  $M_{\rm{ej}}$, $R_{\rm{dec}}$ can even be  smaller than  $R_{\rm{bo}}$. 
For  $R>R_{\rm{dec}}$, the forward shock radius evolves as~\citep{Suzuki:2020qui}
\begin{equation}
\label{Eq: R_sh in dec phase}
    R_{\rm{sh}}(t)=R_{\rm{dec}} \bigg(\frac{t}{t_{\rm{dec}}}\bigg)^{{2}/{3}}\ .
\end{equation}
where we have assumed adiabatic dynamical evolution for the sake of simplicity.
At radii larger than $ R_{\rm{bo}} $, diffusive shock acceleration of the incoming CSM protons takes place.  Following ~\cite{2012ApJ...751...65F,Petropoulou:2016zar}, the proton injection rate for a wind density profile is 
\begin{eqnarray}
	Q_{\rm{p}}(\gamma_{\rm{p}},R) &\equiv& \frac{{\rm{d}}^{2}N_{\rm{p}}}{{\rm{d}}\gamma_{\rm{p}}{\rm{d}}R} = \frac{9 \pi \varepsilon_{\rm{p}} R^{2}_{{\rm{bo}}} n_{{\rm{bo}}}}{8 {\rm{ln}}(\gamma_{p, \rm{max}}/\gamma_{p, \rm{min}})}  \left[\frac{v_{\rm{sh}}(R_{\rm{bo}})}{c}\right]^{2}\\ \nonumber &\times&\bigg(\frac{R}{R_{\rm{bo}}}\bigg)^{2\alpha} \gamma_{\rm{p}}^{-k} H(\gamma_{\rm{p}} - \gamma_{p, \rm{min}}) H(\gamma_{p, \rm{max}} - \gamma_{\rm{p}})\ ,
\end{eqnarray}
where the parameter $\alpha$ dictates the radial dependence of the shock velocity ($v_{\rm{sh}}\propto R^{\alpha}$), it  is $\alpha=-1/7$ in the free expansion phase ($R<R_{\rm{dec}}$) and $\alpha=-1/2$ in the decelerating phase ($R>R_{\rm{dec}}$). The fraction of the shocked thermal energy stored in relativistic protons is $\varepsilon_{\rm{p}} $, while $H(x) = 1$ for $x > 0$ and zero otherwise.  We set the proton spectral index  $ k = 2 $ and the minimum Lorentz factor of the accelerated protons  ${\gamma_{p,\ \rm{min}} = 1}$. The maximum Lorentz factor of  protons ($ \gamma_{p,\ \rm{max}} $) is  obtained by requiring that the acceleration timescale in the Bohm limit, ${ t_{\rm{acc}} \sim 20 \gamma_{\rm{p}}m_{\rm{p}} c^{3} /3 e B v_{\rm{sh}}^{2}}$~\citep{Protheroe:2003vc},  is shorter than the total cooling timescale for protons: $ t_{\rm{acc}} \leq t_{\rm{p,cool}}$.  $ {B = \sqrt{32 \pi \varepsilon_{B} m_{\rm{p}} v_{\rm{sh}}^{2} n_{\rm{CSM}}}} $ is the magnetic field in the post-shock region, whose energy density is a fraction $ \varepsilon_{B} $ of the post-shock thermal energy density $ {U_{\rm{th}} = (9/8) m_{\rm{p}} v^{2}_{\rm{sh}} n_{\rm{CSM}} }$. The latter is obtained by considering the Rankine-Hugoniot jump conditions across a strong non-relativistic shock with compression ratio approximately equal to $4$.

The most relevant energy loss mechanisms for protons are inelastic $ pp $ collisions and the cooling due to  adiabatic expansion of  the shocked shell, hence ${ t^{-1}_{\rm{p,cool}} = t^{-1}_{\rm{pp}} + t^{-1}_{\rm{ad}} }$, with ${ t_{\rm{pp}} = (4 k_{\rm{pp}} \sigma_{\rm{pp}} n_{\rm{CSM}} c) ^{-1}}$, where we assume  constant inelasticity $ k_{\rm{pp}} = 0.5 $ and  energy-dependent cross-section $ \sigma_{\rm{pp}} (E_{\rm{p}})$~\citep{Zyla:2020zbs}.
Following~\cite{Fang:2020bkm}, the adiabatic cooling is  $t_{\rm{ad}}= {\rm{min}}[t_{\rm{dyn}},t_{\rm{cool}}]$, where $t_{\rm{cool}}$ is the typical cooling time of the thermal gas behind the shock and $t_{\rm{dyn}}$ is the dynamical time of the shock. When the shock is radiative, the particle acceleration region is shrank to a characteristic length $\sim v_{\rm{sh}}t_{\rm{cool}}$, limiting the maximum achievable particle energy. The cooling time is  $t_{\rm{cool}}=3 k_{B} T/2 n_{\rm{sh}} \Lambda(T)$~\citep{Franco_1992} where $k_{B}$ is the Boltzmann constant, $n_{\rm{sh}}=4n_{\rm{CSM}}$ is the density of the shocked region, and $\Lambda (T)$ is the cooling function  capturing the physics of radiative cooling. Here $T$ is the gas temperature immediately behind the forward shock front obtained by the Rankine-Hugoniot conditions,  given by:

\begin{equation}
    T=2\frac{(\gamma-1)}{(\gamma+1)^{2}}\frac{\mu m_{\rm{H}} v^{2}_{\rm{sh}}}{k_{B}}\ ,
\end{equation}
where $\gamma=5/3$ is the adiabatic index of the gas. Finally, the cooling function [in units of $\rm{erg}\,\rm{cm}^{3}\,\rm{s}^{-1}$] is given by the following approximation~\citep{1994ApJ...420..268C}:
\begin{equation}
    \Lambda(T)=\begin{cases}
    6.2\times 10^{-19} \, T^{-0.6}\quad 10^{5} < T \lesssim 4.7\times 10^{7}~\rm{K}\\
    2.5\times 10^{-27} \,T^{0.2}\,\,\,\,\,\quad\quad\quad T> 4.7\times 10^{7}~\rm{K}\ .
    \end{cases}
\end{equation}
where line emission dominates at low $T$ and  free-free emission at high $T$.

Relativistic protons in the shocked region may also interact  with the ambient photons via $p\gamma$ interactions. However, in this work we ignore this energy loss channel, consistent with the work of~\cite{Murase:2010cu,Fang:2020bkm}, which show that $p\gamma$ interactions can be safely neglected for a wide range of parameters.

Since we aim to compute the neutrino emission, we track the temporal evolution of the proton distribution in the shocked region between the shock breakout radius $ R_{\rm{bo}} $ and the outer radius $R_{\rm{CSM}}$.

The evolution of the proton distribution is given by~\citep{1997ApJ...490..619S,2012ApJ...751...65F,Petropoulou:2016zar}:
\begin{equation}
	\frac{\partial N_{\rm{p}}(\gamma_{\rm{p}},R)}{\partial R} - \dfrac{\partial }{\partial \gamma_{\rm{p}}}\bigg[\frac{\gamma_{\rm{p}}}{R} N_{\rm{p}}(\gamma_{\rm{p}}, R)\bigg] + \frac{N_{\rm{p}}(\gamma_{\rm{p}}, R)}{v_{\rm{sh}}(R) t_{pp}(R)} = Q_{\rm{p}}(\gamma_{\rm{p}},R)\ ,
\label{eq:Np}	
\end{equation}
where $ N_{\rm{p}} (\gamma_{\rm{p}}, R)$ represents the total number of protons in the shell at a given radius $ R $ with Lorentz factor between $ \gamma_{\rm{p}} $ and $ \gamma_{\rm{p}} + \rm{d}\gamma_{\rm{p}} $. The radius $R$ is related to the  time $t$ measured by an observer at Earth: $t = \tilde{t}(R)(1+z)$, where  we denote with a tilde all parameters in the reference frame of the central compact object hereafter. The second term on the left hand side of Eq.~\ref{eq:Np} takes into account energy losses due to the adiabatic expansion of the SN shell, while  $ pp $ collisions are  treated as an escape term~\citep{1997ApJ...490..619S}. Other energy loss channels for protons are  negligible~\citep{Murase:2010cu}. Furthermore, in Eq.~\ref{eq:Np} the diffusion term has been neglected since the shell is assumed to be homogeneous.

The neutrino production rates, $ Q_{\nu_{i}+\bar{\nu}_{i}} \, [\rm{GeV}^{-1}\rm{cm}^{-1}] $, for muon and electron flavor (anti)neutrinos are given by~\citep{Kelner:2006tc}:
\begin{eqnarray}
\label{eq:Q_nu_mu}
	Q_{\nu_{\mu} + \bar{\nu}_{\mu}}(E_{\nu}, R) &=& \frac{4 n_{\rm{CSM}}(R) m_{\rm{p}} c^{3}}{v_{\rm{sh}}} \int_{0}^{1} dx \frac{\sigma_{\rm{pp}}(E_{\nu}/x)}{x}\\ \nonumber  & & N_{\rm{p}} \bigg(\frac{E_{\nu}}{x m_{\rm{p}} c^{2}}, R\bigg)\bigg(F^{(1)}_{\nu_{\mu}}(E_{\nu}, x) + F^{(2)}_{\nu_{\mu}}(E_{\nu}, x)\bigg)\ , \\
	\label{eq:Q_nu_e}
		Q_{\nu_{e} + \bar{\nu}_{e}}(E_{\nu}, R) &=& \frac{4 n_{\rm{CSM}}(R) m_{\rm{p}} c^{3}}{v_{\rm{sh}}} \int_{0}^{1} dx \frac{\sigma_{\rm{pp}}(E_{\nu}/x)}{x}\\ \nonumber  & & N_{\rm{p}} \bigg(\frac{E_{\nu}}{x m_{\rm{p}} c^{2}}, R\bigg) F_{\nu_{e}}(E_{\nu}, x)\ ,
\end{eqnarray}
where $ x = E_{\nu}/E_{\rm{p}} $. The functions $ F^{(1)}_{\nu_{\mu}} $, $ F^{(2)}_{\nu_{\mu}} $ and $ F_{\nu_{e}} $ follow the definitions in~\cite{Kelner:2006tc}. Equations~\ref{eq:Q_nu_mu} and \ref{eq:Q_nu_e} are valid for $ E_{\rm{p}} > 0.1$~TeV,  corresponding to  the energy range under investigation. 

\subsection{Parameters characteristic of AT 2019fdr}
\label{sec:ingredients}

Numerical simulations aiming to model SLSNe IIn light curves are undergoing, see e.g.~\cite{Dessart:2015xaa,Chatzopoulos2019,Suzuki:2020yhz,Suzuki:2019kny}; however, the exact underlying physics which powers these sources is still uncertain.
In the following, we outline the properties of AT2019fdr useful to model its neutrino emission. 

By relying on existing  data  on AT2019fdr from ZTF \citep{Patterson2019}, ATLAS \citep{ATLAS2018,ATLAS2020} and \textit{Swift} \citep{gehrels04}, we integrate the observed pseudo-bolometric light curve and estimate that the total radiated energy from AT2019fdr is $\tilde{E}_{\rm{rad}}=1.66 \pm 0.01 \times 10^{52}$~erg. To take into account the uncertainties on the radiative efficiency, namely the fraction  of the total energy that is radiated, we consider two characteristic values of the kinetic energy $\tilde{E}_{k}$ of the ejecta: $5\times 10^{52}$~erg and $10^{53}$~erg, which correspond to a radiative efficiency of $\sim 35\%$ and $18\%$, respectively (see \cite{Chevalier:2011ha}, where the total radiated energy is estimated to be $E_{\rm{rad}} = 0.32 E_{\rm{k}}$).

We assume the proton fraction equal to  $ {\varepsilon_{\rm{p}}=0.1 }$~\citep{Murase:2010cu}. This value is   consistent with  simulations of particle acceleration and magnetic field amplification at non-relativistic quasi-parallel shocks~\citep{Caprioli:2013dca}. A discussion on  the impact of  different values of ${\varepsilon_{\rm{p}}}$ on the expected neutrino event rate is reported in Sec.~\ref{sec:discussion}. The fraction of the post-shock internal energy that goes into amplification of the magnetic field is instead assumed to be $ \varepsilon_{B}=3\times 10^{-4} $~\citep{Petropoulou:2017ymv}. 

The wind velocity is considered to be $ v_{w} = 100$~km s$^{-1} $~\citep{Moriya:2014cua}. The average mass loss rate is given by~\citep{Suzuki:2020yhz}:
\begin{equation}
	\dot{M}=0.3\ {M}_{\odot}\ {\rm{yr}}^{-1} \bigg(\frac{M_{\rm{CSM}}}{10\ M_{\odot}}\bigg)\bigg(\frac{R_{\rm{CSM}}}{10^{16}\ \rm{cm}}\bigg)^{-1}\bigg(\frac{v_{w}}{100\ \rm{km\ s}^{-1}}\bigg)\ ,
\end{equation}
where $ M_{\rm{CSM}} $ is the CSM mass contained within a shell of radius $R_{\rm{CSM}}$.

By fitting a basic polynomial to the bolometric light curve and available non-detections of AT2019fdr to extrapolate beyond the detection limits of ZTF, we estimate that its rise time  (i.e., the time during which the luminosity reaches peak value,  see Fig.\ref{Fig:LC_SN}) is $t_{\rm{rise}}\sim 98$~days. In addition, in order to link $t_{\rm{rise}}$ to the other model parameters characteristic of AT2019fdr,  we rely on the following relation~\citep{Suzuki:2020qui}:
\begin{equation}
\label{Eq: t_diff}
  t_{\rm{rise}} \simeq t_{\rm{diff}}(t)=\frac{(R_{\rm{ph}}-R) \tau_{\rm{T}}(R)}{c}\ ,
\end{equation}
where the diffusion time is the time required for the radiation to travel from $R$ to  $R_{\rm{ph}}$~\footnote{The photospheric radius is obtained by considering  ${\tau_{T}(R_{\rm{ph}}) = 1}$.}, and $\tau_{\rm{T}}(R)$ is the optical depth of the CSM at radius $R$. The rise time is expected to increase as a function of $M_{\rm{CSM}}$, since a massive and dense CSM prolong the photon diffusion timescale. Yet, in order to predict the correct behavior of $t_{\rm{rise}}$, one should take into account the effect of the variation of all the parameters: $E_{\rm{k}}$, $M_{\rm{ej}}$, $M_{\rm{CSM}}$, and $R_{\rm{CSM}}$.

The exact values of $ M_{\rm{ej}} $, $ M_{\rm{CSM}} $, and $ R_{\rm{CSM}} $ are highly uncertain for AT2019fdr and degeneracies may be at play when interpreting the AT2019fdr light curve. The reprocessing of information from the explosion by interaction with the CSM masks the properties of the SLSN explosion underneath it. Although the CSM density can be estimated in several ways, e.g.~from the strength of the $H$--$\alpha$ 
line~\citep{Taddia2013} or through X-ray and radio observations~\citep{Chandra2018}, AT2019fdr lacks the necessary time series multi-wavelength and spectroscopic data required to constrain it.
Hence,  we consider ranges of variability for the most uncertain parameters: $M_{\rm{ej}} \in [5, 150]\ {M}_{\odot}$, $M_{\rm{CSM}} \in [5, 150]\ {M}_{\odot}$, and $ R_{\rm{CSM}} \in [2, 4] \times 10^{16}$~cm. Out of these, we only take into account  the ones in agreement with the measured $t_{\rm{rise}} $ (allowing for an uncertainty of $50\%$) and requiring that the production of the neutrinos observed by the IceCube Observatory at $\sim 394$~days after the breakout takes place inside the CSM, namely $t(R_{\rm{CSM}})-t(R_{\rm{bo}})\gtrsim 394$~days. See  Appendix~\ref{Appendix: parameter space} for more details.
A summary of the default values for the parameters considered for AT2019fdr is reported in Table~\ref{table:parameters}.
\begin{table}[ht]
	\caption{\label{table:parameters}Benchmark values for the parameters characteristic of AT2019fdr. For the most uncertain ones, we consider a range of variability.}
	\begin{center}
\hspace{-0.9cm}
	\begin{tabular}{ccc}
		\toprule
		{\it Parameter} & {\it Symbol} & {\it Default value}\\
		\toprule
		Radiated energy & $\tilde{E}_{\rm rad}$ & $1.66 \times 10^{52}$~erg  \\
		Radiative efficiency& $\eta$ & $18$--$35 \%$\\
	   Rise time & $t_{\rm rise}$ & $98$~days  \\
		Redshift& $z$& $0.2666$\\
		Declination & $\delta$& $26.85$~deg\\
		Right ascension & $\alpha$& $257.28$~deg\\
		Accelerated proton energy fraction& $ \varepsilon_{\rm{p}} $& $0.1$\\
		Magnetic energy density  fraction & $ \varepsilon_{B} $&$ 3\times 10^{-4} $\\
		Proton spectral index& $k$ & $2$\\
		Wind velocity& $ v_{w} $& $ 100 \,\rm{km\, s}^{-1}$\\
		Ejecta density slope & $ s $& $ 10 $\\
		Ejecta mass& $ M_{\rm{ej}} $& $ 5$--$150\ {M}_{\odot} $\\
		CSM mass& $ M_{\rm{CSM}} $& $ 5$--$150\ {M}_{\odot} $\\
		CSM radius & $ R_{\rm{CSM}} $& $ (2$--$4) \times 10^{16} $~cm\\
		\toprule
	\end{tabular}
	\end{center}
\end{table}

\subsection{Neutrino flux and event rate at Earth}
\label{sec:nurateflux}

The neutrino and antineutrino flux ($F_{\nu_{\alpha} +\bar{\nu}_{\alpha}}$ with $\alpha=e, \mu, \tau$)  at Earth from a SN at redshift $z$ and as a function of time in the observer frame is [$\mathrm{GeV}^{-1} \mathrm{s}^{-1}\mathrm{cm}^{-2}$]:
\begin{equation}
\label{eq:F}
	F_{\nu_{\alpha}+\bar{\nu}_{\alpha}}(E_{\nu}, t) = \frac{(1 + z)^2}{4 \pi d^{2}_{L}(z)} v_{\rm{sh}}(t)\underset{\beta}{\sum} P_{\nu_{\beta} \rightarrow \nu_{\alpha}}Q_{\nu_{\beta} +\bar{\nu}_{\beta}}(E_{\nu_{\alpha}}(1 + z),R(t))\ ,
\end{equation}
where $Q_{\nu_{\beta} +\bar{\nu}_{\beta}}$ is defined as in Eqs.~\ref{eq:Q_nu_mu} and \ref{eq:Q_nu_e}. Neutrinos change their flavor while propagating, hence the flavor transition probabilities are given by~\citep{Anchordoqui:2013dnh}:
\begin{eqnarray}
	P_{\nu_{e}\rightarrow\nu_{\mu}} &=& P_{\nu_{\mu}\rightarrow\nu_{e}} = P_{\nu_{e}\rightarrow\nu_{\tau}} = \frac{1}{4}\sin^{2}2\theta_{12}\ ,\\
	P_{\nu_{\mu}\rightarrow\nu_{\mu}}&=& P_{\nu_{\mu}\rightarrow\nu_{\tau}} = \frac{1}{8}(4 - \sin^{2}2\theta_{12})\ ,\\
	P_{\nu_{e}\rightarrow\nu_{e}}&=& 1-\frac{1}{2}\sin^{2}2\theta_{12}\ ,
\end{eqnarray}
with $\theta_{12} \simeq 33.5$~deg~\citep{Esteban:2020cvm}, and  $P_{\nu_{\beta}\rightarrow \nu_{\alpha}} = P_{\bar{\nu}_{\beta}\rightarrow \bar{\nu}_{\alpha}}$. The  luminosity distance   $ d_{L}(z) $ is  defined in a flat $\Lambda$CDM  cosmology as
\begin{equation}
	\label{luminosity_distance}
	d_{L}(z) = (1+z) \frac{c}{H_{0}} \int_{0}^{z} \frac{dz^\prime}{\sqrt{\Omega_{\Lambda}+\Omega_{M}(1+z^\prime)^3}}\ ,
\end{equation}
where $\Omega_{M} = 0.315$, $\Omega_{\Lambda} = 0.685$ and the Hubble constant is ${H_{0} = 67.4}$~km s$^{-1}$ Mpc$^{-1}$~\citep{Aghanim:2018eyx}.

The  neutrino fluence [$\mathrm{GeV}^{-1} \mathrm{cm}^{-2}$] is calculated using
\begin{equation}
	\label{Eq: neutrino fluence}
	\Phi_{\nu_{\alpha}+\bar{\nu}_{\alpha}}(E_{\nu}) = \int_{t_{\rm{bo}}}^{t_{\rm{bo}}+394} F_{\nu_{\alpha}+\bar{\nu}_{\alpha}}(E_{\nu}, t) dt\ ,
\end{equation}
with $ t_{\rm{bo}} = t(R_{\rm{bo}})$ and the time integral being restricted to  $394$~days.

Finally, the  event rate of muon neutrinos and antineutrinos expected at the IceCube Neutrino Observatory is 
\begin{equation}
	\label{eq: neutrino event number}
	\dot{N}_{\nu_{\mu}+\bar{\nu}_{\mu}}(t)= \int_{E_{\nu,\ \rm{min}}}^{E_{\nu,\ \rm{max}}} dE_{\nu} A_{\rm{eff}}(E_{\nu}, \delta)  F_{\nu_{\mu}+\bar{\nu}_{\mu}}(E_{\nu}, t)
\end{equation}
where $ A_{\rm{eff}}(E_{\nu},\delta)$ is the detector effective area~\citep{IceCube:2021xar}. The minimum neutrino energy is  $ E_{\nu,\ \rm{min}}=100 $~GeV  for the  declination of interest~\citep{IceCube:2021xar}, and $F_{\nu_{\mu}+\bar{\nu}_{\mu}}(E_{\nu}, t)$ has been introduced in Eq.~\ref{eq:F}.  In the following, we work under the assumption of perfect discrimination between astrophysical and atmospheric neutrinos;  see Sec.~\ref{sec:discussion} for a discussion on the expected event rate if the event sample should be contaminated by atmospheric neutrinos in the energy region below $100$~TeV. The maximum neutrino energy $E_{\nu,\rm{max}}$ is related to the maximum proton energy: $E_{\nu,\rm{max}} = x E_{p,\rm{max}}$. 

The total number of muon neutrinos and antineutrinos is computed over the temporal interval of $394$ days: 
\begin{equation}
\label{eq:Ntot}
N_{\nu_{\mu}+\bar{\nu}_{\mu}}= \int_{t_{\rm{bo}}}^{t_{\rm{bo}}+394}dt\ \dot{N}_{\nu_{\mu}+\bar{\nu}_{\mu}}(t)\ .
\end{equation}

\section{Forecast of the  neutrino signal}
\label{sec:results}
In this section, we present the results on the  neutrino signal expected from AT2019fdr. First, we discuss the neutrino spectral energy distribution and the event rate  expected in the IceCube Neutrino Observatory. We then  investigate  the dependence of the expected signal on the uncertainties of  the SLSN IIn model.

\subsection{Energy fluence and temporal evolution of the neutrino event rate}
\begin{figure}
	\centering
	\includegraphics[width=0.45\textwidth]{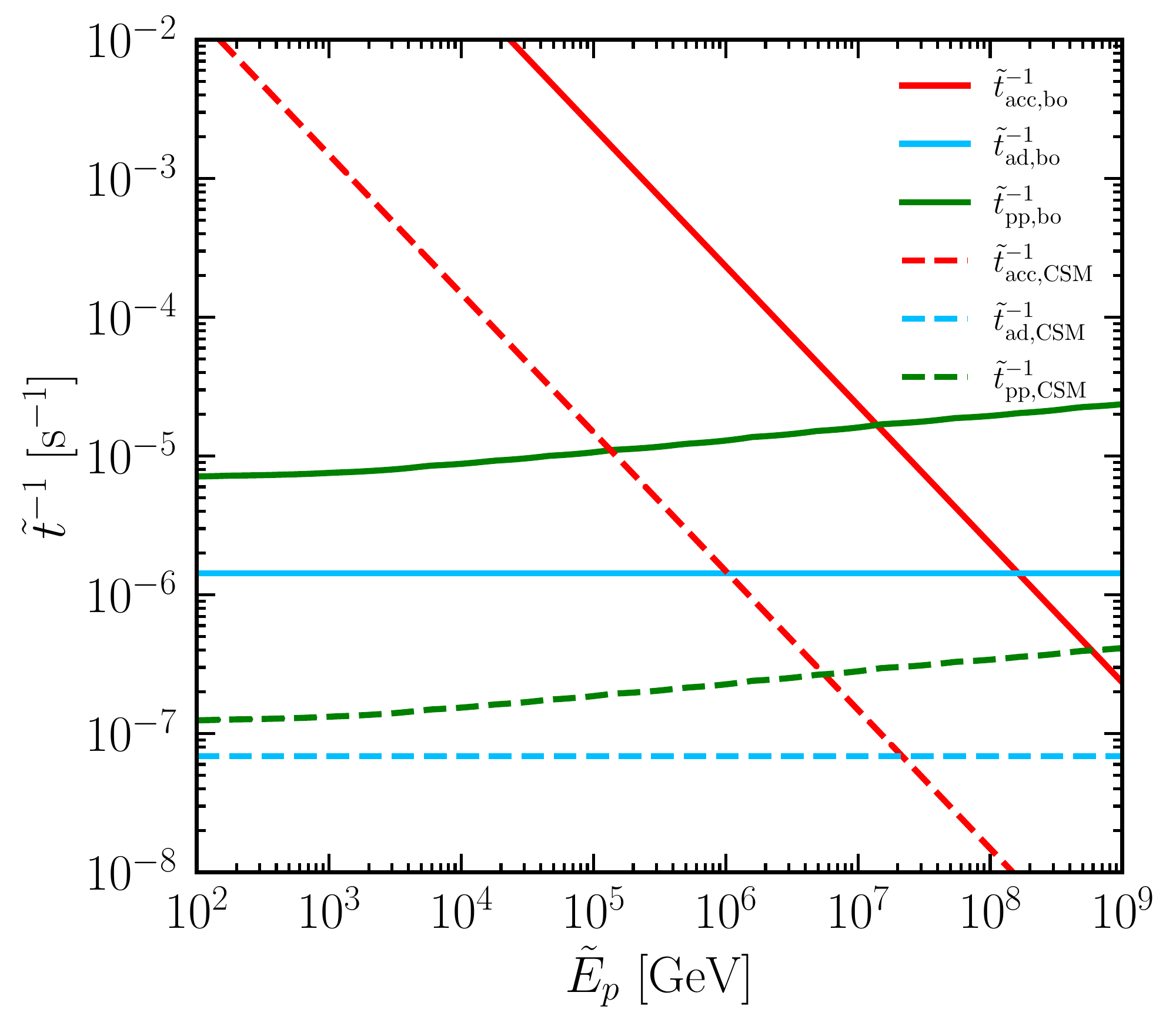}
	\caption{Inverse cooling  and acceleration timescales for protons at the breakout radius ($R_{\rm{bo}}$, solid lines) and at the outer edge $R_{\rm{CSM}}$ (dashed lines) as functions of the proton
    energy in rest frame for the SLSN configuration with  $(\tilde{E}_{\rm{k}},R_{{\rm{CSM}}},  M_{\rm{ej}}, M_{\rm{CSM}}) = (10^{53}\,{\rm{erg}}, 4 \times 10^{16}\ {\rm{cm}}, 6\ M_{\odot}, 49\ M_{\odot})$. The acceleration timescale, $pp$  and adiabatic cooling timescales are represented in red, green and light blue, respectively. Protons are strongly cooled by $pp$ energy losses for all the SLSN parameter configurations considered in this work.}
	\label{Fig:cooling times}
\end{figure}

Before  focuing on the energy fluence and event rate of the detectable neutrino signal, we explore the characteristic cooling times of protons and the acceleration timescale characteristic of AT2019fdr, introduced in Sec.~\ref{sec:energy_distributions}. In order to give an idea of the variation of the cooling and acceleration timescales across the SLSN shell, Fig.~\ref{Fig:cooling times} shows  the proton cooling times as a function of the proton energy in the reference frame of the central compact object and at the representative radii $R_{\rm{bo}}$ and $R_{\rm{CSM}}$ for the  SLSN configuration with $(\tilde{E}_{\rm{k}}, R_{{\rm{CSM}}},  M_{\rm{ej}}, M_{\rm{CSM}}) = (10^{53}\,{\rm{erg}}, 4 \times 10^{16}\ {{\rm{cm}}}, 6\,M_{\odot}, 49\,M_{\odot})$. 
As discussed in the following, this SLSN configuration leads to the most optimistic scenario for  neutrino production.

Proton-proton collisions are responsible for the  dominant energy loss channel. Even though Fig.~\ref{Fig:cooling times} represents the characteristic cooling times for one specific SLSN configuration, the hierarchy between $pp$ and adiabatic losses is representative of all SLSN configurations considered in this work 
(lower $\tilde{E}_{\rm{k}}$ and $R_{\rm{CSM}}$ larger than the ones adopted here would lead to scenarios with adiabatic energy losses being dominant over  $pp$ ones). 

The evolution of $E_{\rm{p,max}}$ depends on the specific choice of parameters $\tilde{E}_{\rm{k}}, R_{\rm{CSM}}, M_{\rm{ej}}$, and $R_{\rm{CSM}}$,  determining whether $R_{\rm{bo}} \lessgtr R_{\rm{dec}}$. For the typical values of $\tilde{E}_{\rm{k}}$ and $R_{\rm{CSM}}$ considered in this work, the condition $t_{\rm{pp}}< t_{\rm{ad}}$ is always fulfilled, and $E_{\rm{p,max}}$  increases as a  function of $R$ up to $R_{\rm{dec}}$, and decreases otherwise. In fact, by using Eqs.~\ref{eq:n_CSM}, \ref{Eq: Rsh in free phase} and \ref{Eq: R_sh in dec phase}, we find: 
\begin{equation}
\label{eq:Ep_max}
    E_{\rm{p, max}} = \frac{e B v_{\rm{sh}}^{2}}{24 c^{2} k_{\rm{pp}} \sigma_{\rm{pp}} n_{\rm{CSM}}}\propto \, \begin{cases}
     M_{\rm{ej}}^{-{15}/{14}}M_{\rm{CSM}}^{-{13}/{14}}\, R^{{4}/{7}}\quad &R<R_{\rm{dec}}\\
     M_{\rm{CSM}}^{-2} \,R^{-{1}/{2}}\quad &R\geq R_{\rm{dec}} \ .
    \end{cases}
\end{equation}
Appendix~\ref{Appendix: maximum_proton_energy} provides more details on the scaling of $E_{\rm{p, max}}$ as a function of the SLSN model parameters.

\begin{figure}
	\centering
	\includegraphics[width=0.45\textwidth]{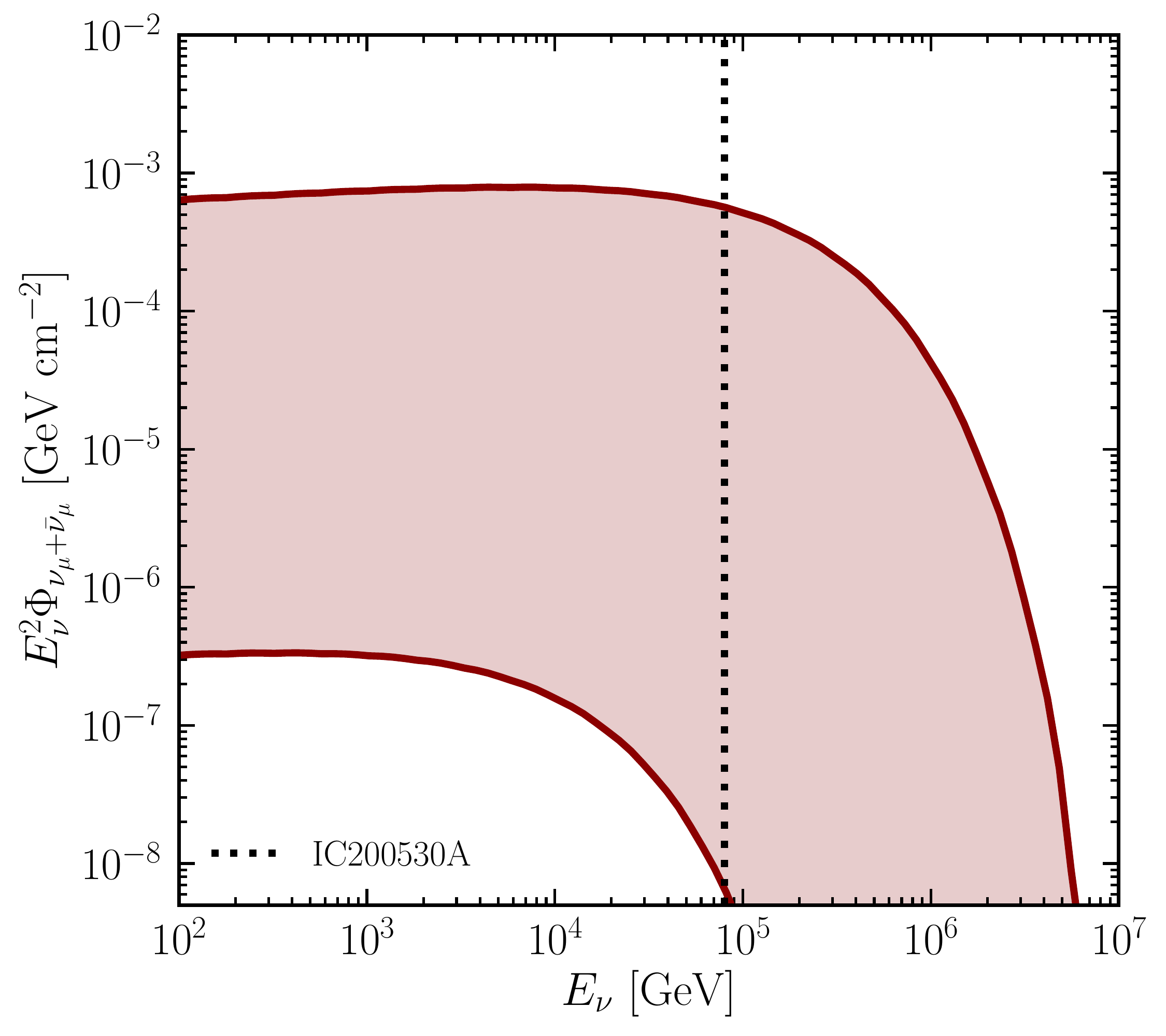}
	\caption{Muon neutrino and antineutrino fluence from AT2019fdr as a function of the neutrino energy. The reconstructed neutrino energy ($ E_{\nu}\sim 80 $~TeV) for IC200530 is marked by a black dotted vertical line.  The band encloses the uncertainties on the parameters characterizing AT2019fdr, see Table~\ref{table:parameters}. In the proximity of  the energy of interest for the interpretation of IC200530, the fluence can vary up to a factor  $\mathcal{O}(10^{5})$ in magnitude. Within the allowed parameter space, the  lowest fluence is foreseen for  configurations with large  $R_{\rm{CSM}}$, low  $M_{\rm{CSM}}$ and high  $M_{\rm{ej}}$. The largest neutrino fluence is instead obtained for  intermediate values of  $M_{\rm{CSM}}$ and  low $M_{\rm{ej}}$, which moreover allow a higher proton energy cutoff. }
	\label{Fig:Fluence}
\end{figure}

The  muon neutrino and antineutrino fluence, defined as in Eq.~\ref{Eq: neutrino fluence}, is shown in Fig.~\ref{Fig:Fluence} as a function of the neutrino energy. The band takes into account the uncertainties on the parameters characterizing AT2019fdr (see Sec.~\ref{sec:ingredients}) and is defined by the parameter configurations leading to the  highest and lowest neutrino fluence.
Note that, for the SLSN parameters adopted in this work, the synchrotron cooling of charged pions and muons produced via $pp$ interactions is  negligible. In fact, the typical energies for which this energy loss becomes relevant  are at least three orders of magnitude larger than the maximum achievable proton energies. Therefore, the neutrino spectra are not affected by the cooling of mesons. 

Given our selection criterion (i.e., the observation of IC200530 about $394$~days after the shock breakout and the constraints on  the rising time of the light curve of AT2019fdr), the scenarios with the lowest fluence are the ones corresponding to configurations with large $R_{\rm{CSM}}$, low $M_{\rm{CSM}}$ and high $M_{\rm{ej}}$. On the other hand, given the reduced parameter space allowed for low $R_{\rm{CSM}}$ (see Appendix~\ref{Appendix: parameter space}), 
the most optimistic scenario corresponds to the highest $R_{\rm{CSM}}$, the lowest accessible $M_{\rm{ej}}$, and intermediate values of $M_{\rm{CSM}}$ ($M_{\rm{CSM}}\simeq 30$--$50 M_{\odot}$).
We refer the reader to Sec.~\ref{sec:param_dep} for a discussion on the dependence of the neutrino fluence from the SLSN characteristic parameters.

The reconstructed neutrino energy for the IC200530 neutrino event is marked with a dotted vertical line and it falls in the same energy range as  the predicted fluence.  One can see that, around the reconstructed energy of IC200530, the fluence can vary up to $ \mathcal{O}(10^5) $  in magnitude. However, it is worth noting that the reconstructed energy carries an intrinsic uncertainty and may differ from the real energy of the detected neutrino, nevertheless we show it in order to guide the eye.

\begin{figure}
	\centering
	\includegraphics[width=0.45\textwidth]{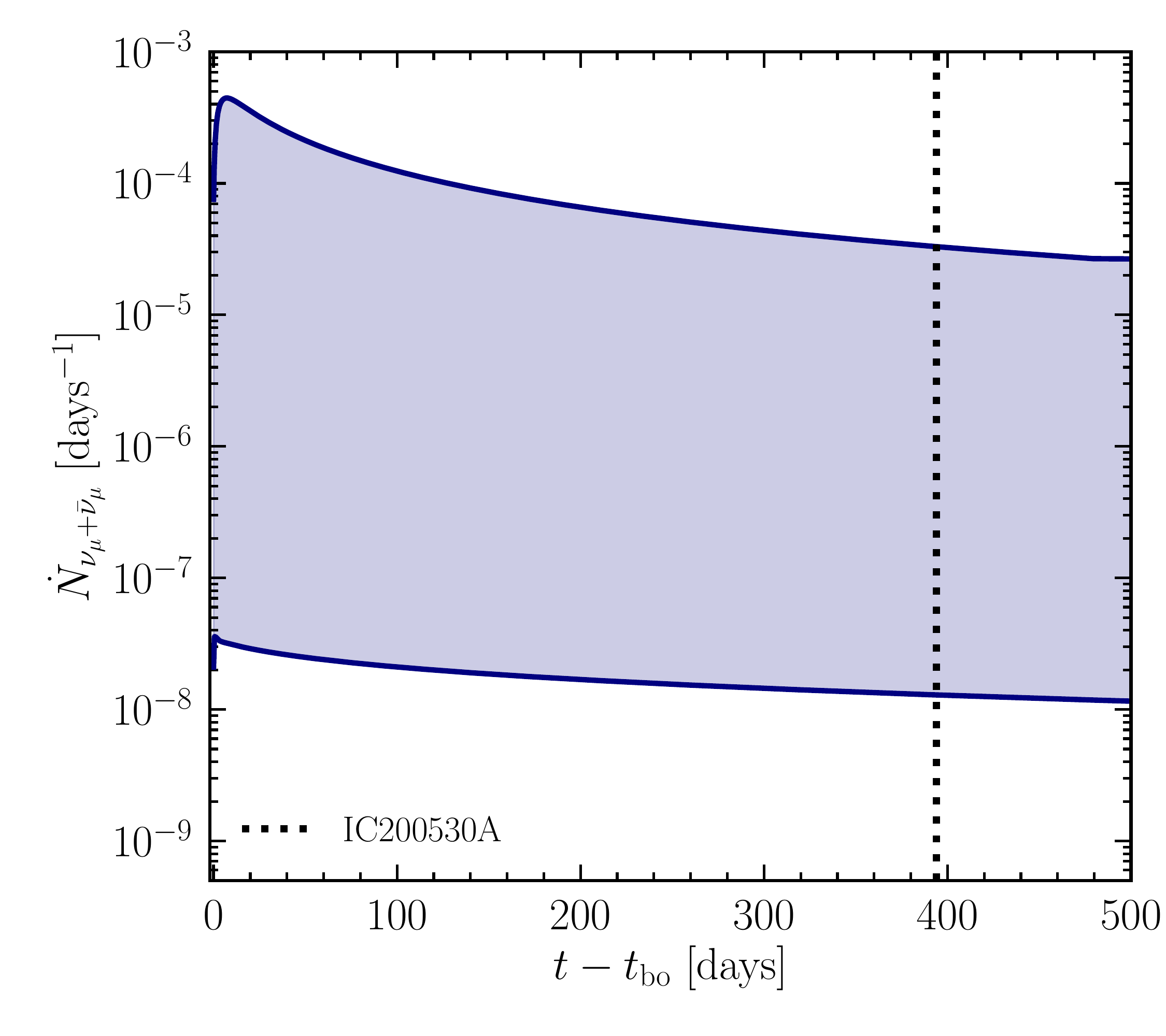}
	\caption{Muon neutrino and antineutrino event rate expected at the IceCube Neutrino Observatory from AT2019fdr as a function of the time after the shock breakout. The band marks the uncertainty on the neutrino event rate due to the SLSN model parameters, see Table~\ref{table:parameters}. The event rate increases rapidly at early times. After  peak,  the event rates for  the SLSN scenarios representing the edges of the envelope decline because of the dominant decreasing trend of $v_{\rm{sh}}$ as a function of time. In some intermediate scenarios, the increasing trend of $E_{\rm{p,max}}$ and shallow decrease of $v_{\rm{sh}}$ can be compensated, providing an increasing event rate at the moment of the detection. The neutrino event IC200530 has been observed $ \sim 394 $ days after  $t_{\rm{bo}}$ as indicated by the dotted vertical line.  In  the  proximity  of  the detection day,  the  event rate can vary up to a factor $\mathcal{O}(10^{3})$ in magnitude.}
	\label{Fig:N_numu_rate_band}
\end{figure}

The muon neutrino and antineutrino event rate expected in IceCube (Eq.~\ref{eq: neutrino event number}) is shown in Fig.~\ref{Fig:N_numu_rate_band} as a function of time. The  band in Fig.~\ref{Fig:N_numu_rate_band} takes into account the uncertainties on the characteristic quantities of AT2019fdr summarized in Table~\ref{table:parameters}. 
For all SLSN cases within the envelope in Fig.~\ref{Fig:N_numu_rate_band}, the event rate increases rapidly at early times. After the peak, depending on whether $R_{\rm{dec}}> R_{\rm{bo}}$ or $R_{\rm{dec}}< R_{\rm{bo}}$, the neutrino event rate has a steeper or  shallower decay. These two different trends are related to the evolution of the shock velocity and the maximum proton energy $E_{\rm{p,max}}$. Indeed,  $E_{\rm{p,max}}$ increases up to $R_{\rm{dec}}$ as $t$ increases and declines later. Since the detector effective area $A_{\rm{eff}}$ increases as a function of $E_{\nu}$~\citep{IceCube:2021xar} and the decline of $v_{\rm{sh}}$ for $R_{\rm{bo}}<R<R_{\rm{dec}}$ is shallow, a compensation effect can arise among the two quantities; hence,  the drop of the  $\dot{N}_{\nu_{\mu}+\bar{\nu}_{\mu}}$ curve can be  slow. Viceversa, when both $E_{\rm{p,max}}$ and $v_{\rm{sh}}$ decrease, the event rate drops faster.  
Around the day of detection of IC200530 (marked by a vertical dotted line), the neutrino event rate is expected to vary between $[1.3\times 10^{-8}, 3.3\times 10^{-5}]$~days$^{-1}$.

It is important to note that only a sub-sample of the SLSN parameter set reported in Table~\ref{table:parameters} allows us to obtain a neutrino signal compatible with our observational constraints. For example, none of the SLSN scenarios with $\tilde{E}_{\rm{k}}=10^{53}$~erg and $ {R_{\rm{CSM}}=2\times 10^{16} }$~cm passes our selection criteria, since the shock crosses the CSM envelope in a time shorter than $ 394 $ days.

\subsection{Dependence of the neutrino signal on the   parameters  of AT2019fdr}
\label{sec:param_dep}

\begin{figure*}
	\centering
	\includegraphics[width=0.45\textwidth]{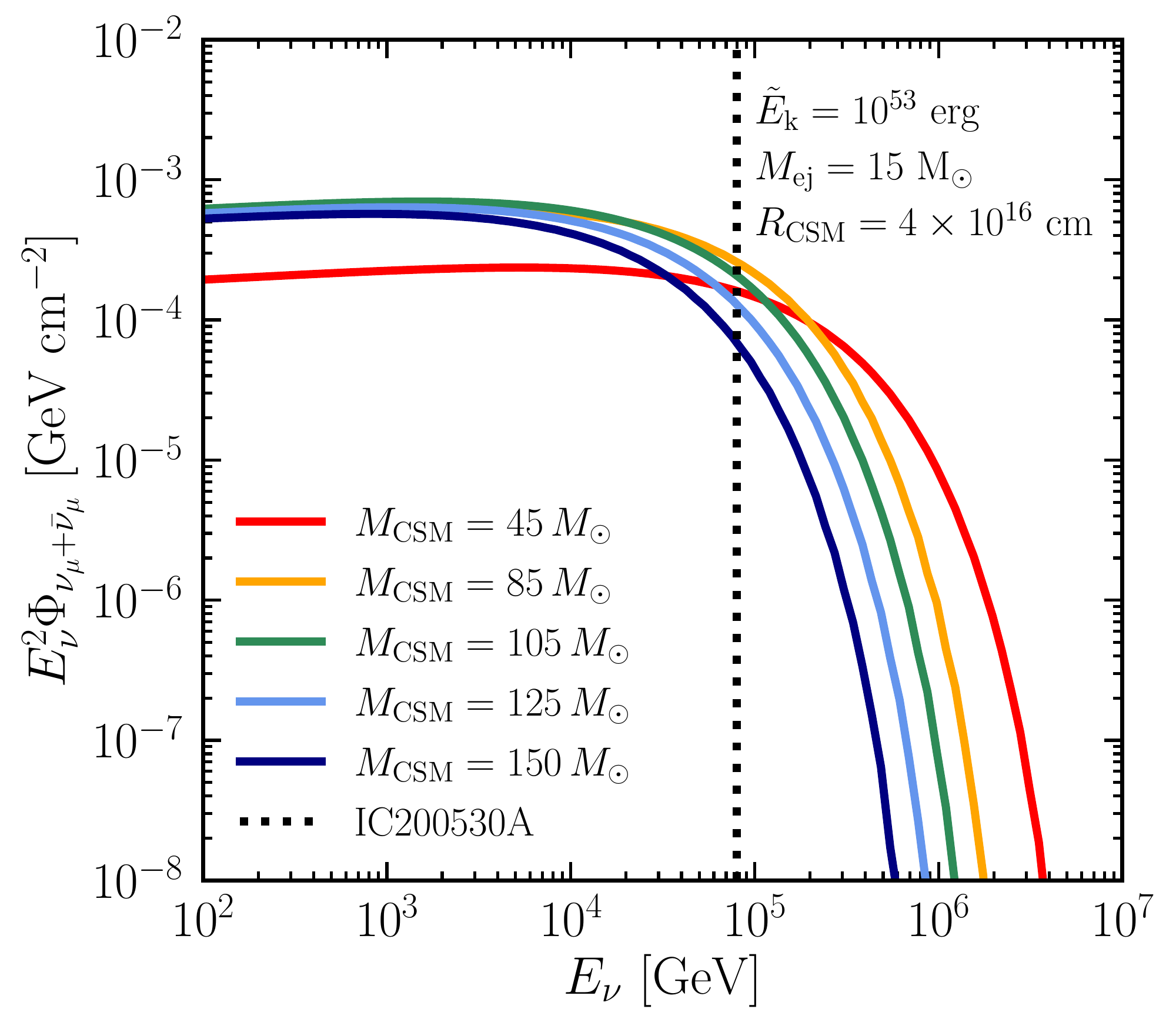}
	\includegraphics[width=0.45\textwidth]{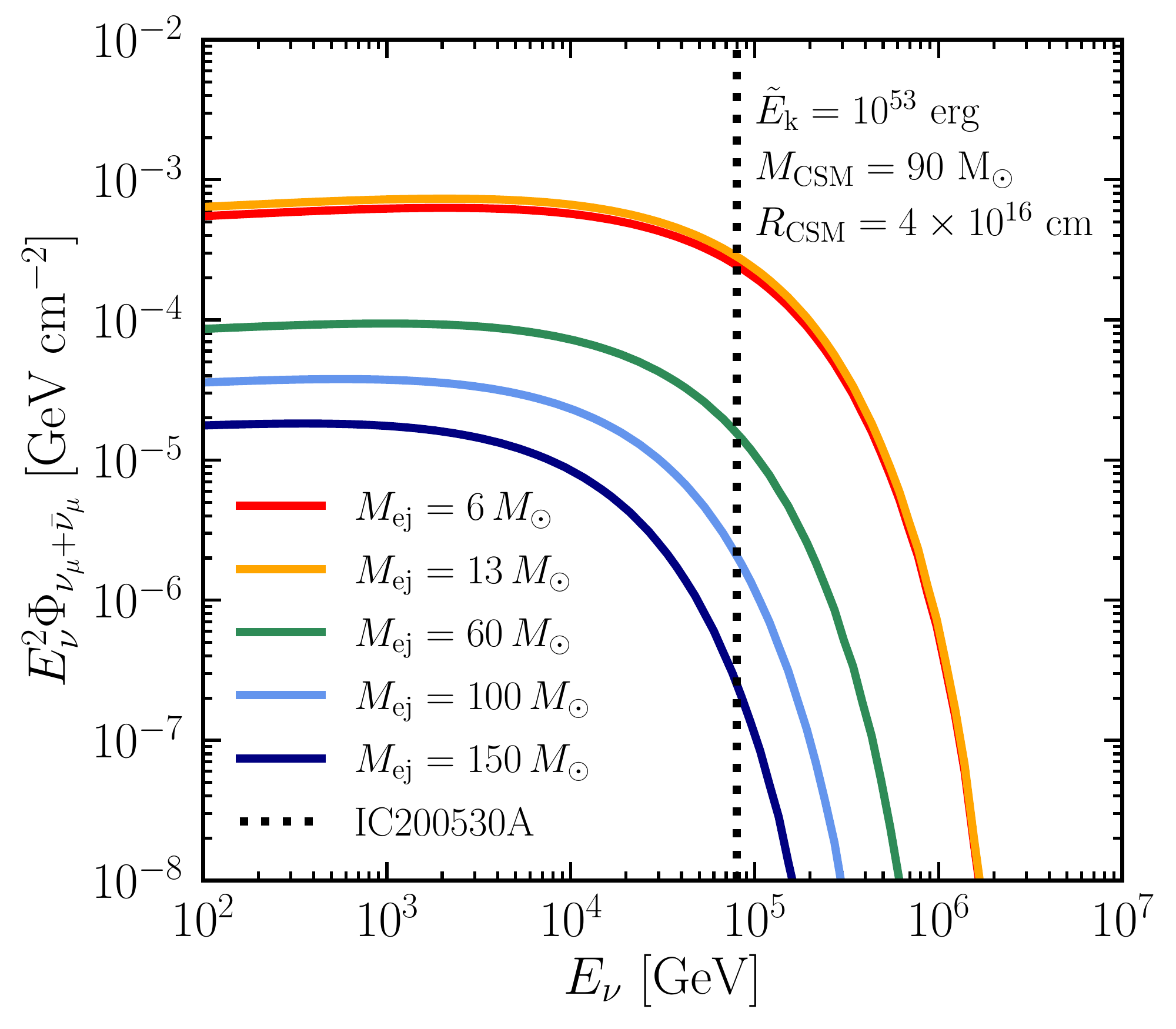}
	\caption{\textit{Left panel}: Muon neutrino and antineutrino fluence for AT2019fdr as a function of the neutrino energy for fixed $ R_{\rm{CSM}} $, $ M_{\rm{ej}} $, $\tilde{E}_{\rm{k}}=10^{53}$~erg and variable $ M_{\rm{CSM}} $. The fluence increases as  $ M_{\rm{CSM}} $ increases up to $M_{\rm{CSM}}=85\,M_{\odot}$, where one transitions into the regime with $R_{\rm{bo}}>R_{\rm{dec}}$. Then a slow decrease in amplitude is observed.
	Furthermore, a slight shift of the neutrino cutoff energy towards lower energies occurs as $ M_{\rm{CSM}} $ increases because of the enhanced $ pp $ energy loss which  prevents particles from being accelerated to higher energies.  \textit{Right panel}: Same as in the left panel, but for fixed $ M_{\rm{CSM}} $ and variable $ M_{\rm{ej}} $. The fluence increases as $ M_{\rm{ej}}$ decreases. This trend is inverted for $M_{\rm{ej}}\lesssim 13\,M_{\odot}$, since  $R_{\rm{bo}}>R_{\rm{dec}}$, and thus the overall shock velocity becomes lower.
	}
	\label{Fig: Fluence R3 Mej=20 and R3 MCSM=30}
\end{figure*}

In order to better explore  the dependence of the neutrino signal expected in IceCube on $ M_{\rm{ej}} $ and $ M_{\rm{CSM}} $,  for $\tilde{E}_{\rm{k}}=10^{53}$~erg, first we investigate  the neutrino fluence as a function of $ M_{\rm{CSM}} $ for fixed $ R_{\rm{CSM}}$ and $ M_{\rm{ej}} $ and then we fix $ M_{\rm{CSM}} $ and vary $ M_{\rm{ej}}$. 
The choice of $M_{\rm{CSM}}$ and $M_{\rm{ej}}$ is guided by the SLSN configurations that better highlight the changes in the neutrino fluence for  $R_{\rm{bo}}\lessgtr R_{\rm{dec}}$. From the  panel on the left in Fig.~\ref{Fig: Fluence R3 Mej=20 and R3 MCSM=30}, we see that the fluence increases as  $ M_{\rm{CSM}} $ increases up to $M_{\rm{CSM}}=85 \,M_{\odot}$. For larger $M_{\rm{CSM}}$,  $R_{\rm{bo}}>R_{\rm{dec}}$, and therefore a turnover with a slow drop can be observed.
Furthermore, a slight shift of the neutrino cutoff energy towards lower energies is visible as  $ M_{\rm{CSM}} $ increases. The latter is due to the enhanced $ pp $ energy loss determined by the larger density as well as the smaller $v_{\rm{sh}}$, which prevent particles from being accelerated to higher energies (see Eq.\ref{eq:Ep_max}).

In the right panel of Fig.~\ref{Fig: Fluence R3 Mej=20 and R3 MCSM=30}, we observe an enhancement of the fluence as $ M_{\rm{ej}}$ decreases. Nevertheless, this trend is inverted for $M_{\rm{ej}}\lesssim 13\,M_{\odot}$, representative of the regime with $R_{\rm{bo}}>R_{\rm{dec}}$, where the lower $v_{\rm{sh}}$ is responsible for  a slight decrease in the neutrino production, together with a shift of the neutrino energy cutoff  to lower energies.

\begin{figure*}
	\begin{center}
		\includegraphics[width=0.45\textwidth]{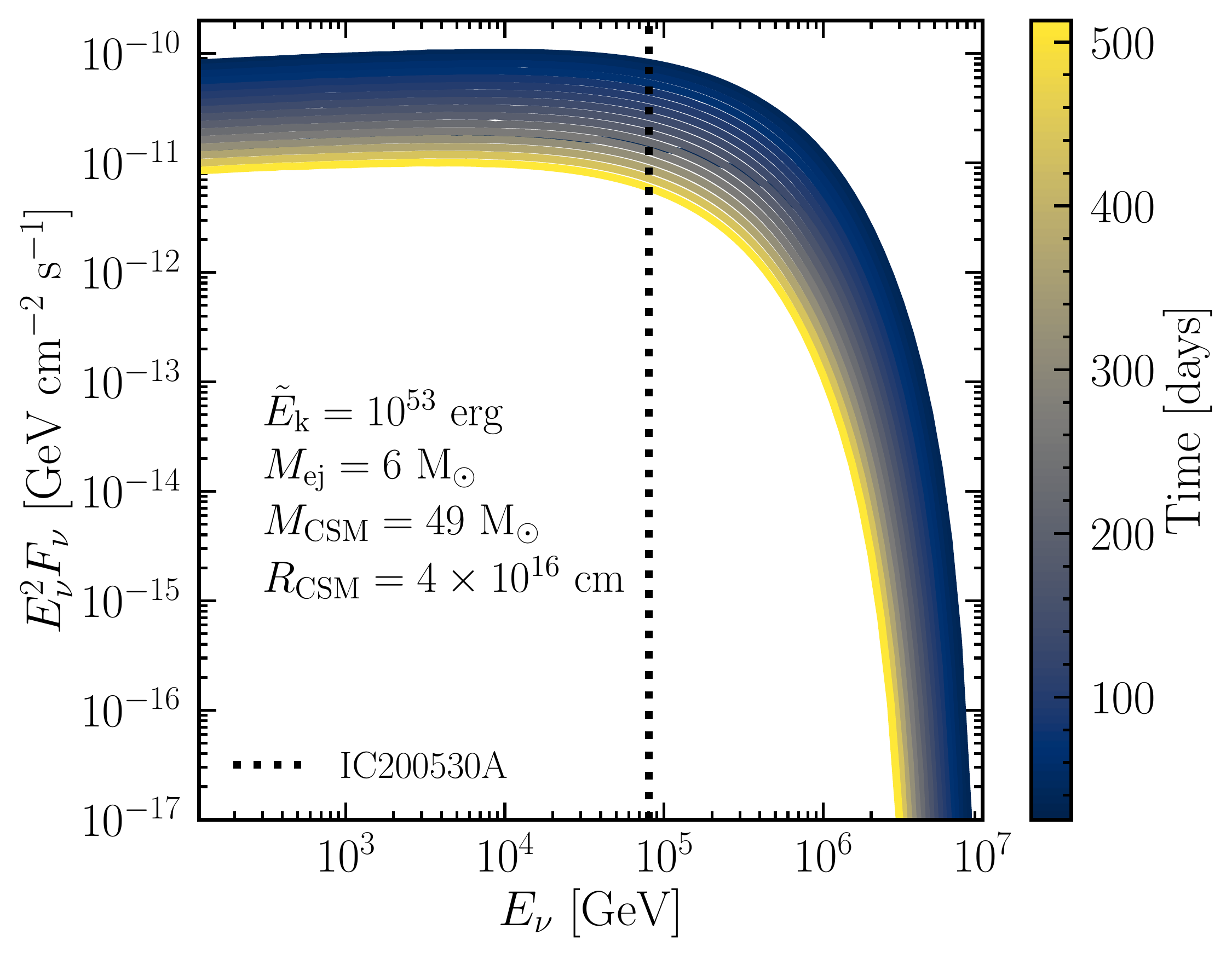}
		\includegraphics[width=0.45\textwidth]{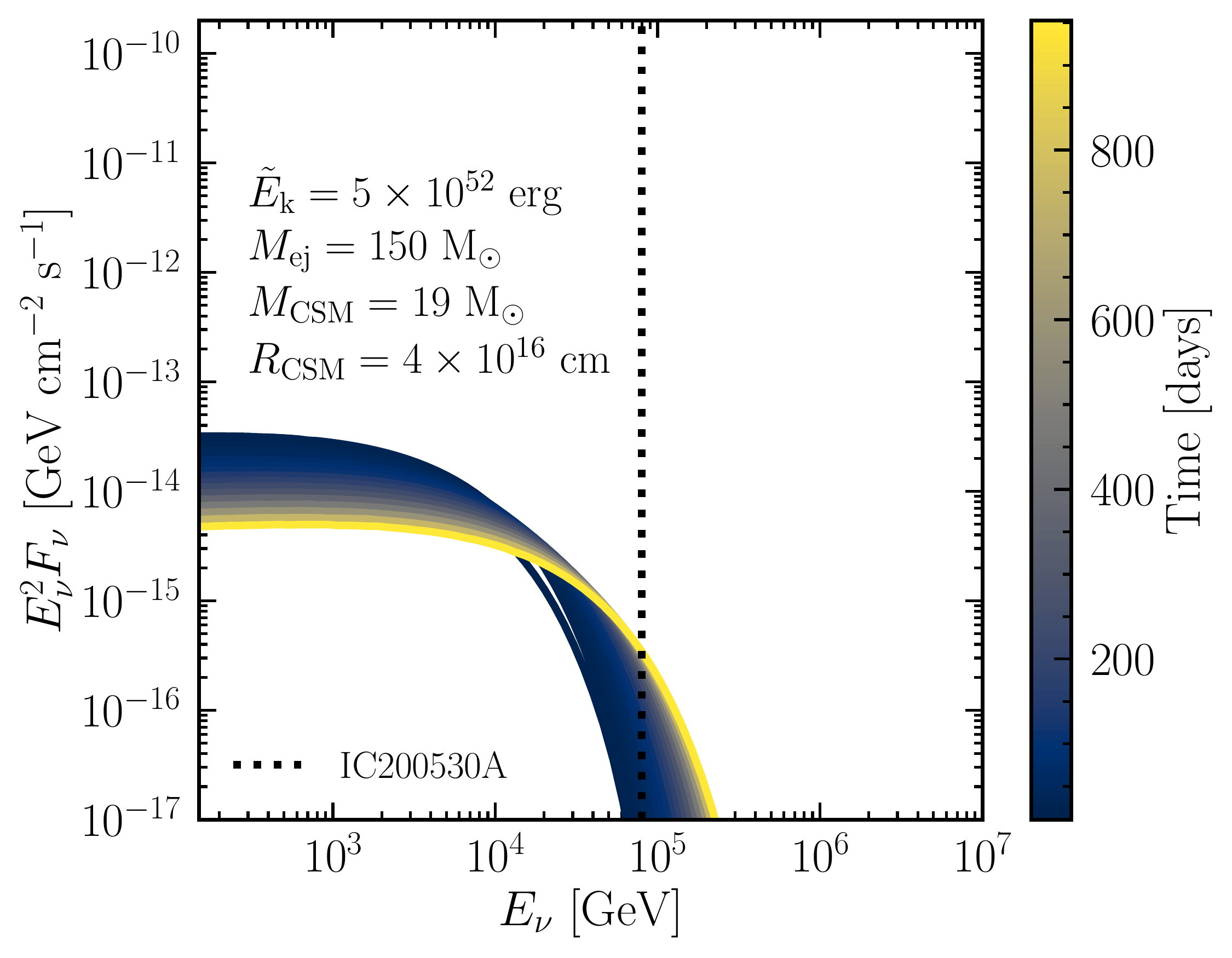}
	\end{center}
	\caption{Temporal evolution of muon neutrino and antineutrino flux from AT2019fdr as a function of the neutrino energy for the most optimistic, $(\tilde{E}_{\rm{k}},R_{\rm{CSM}}, M_{\rm{ej}},  M_{\rm{CSM}}) = (10^{53}\, {\rm{erg}}, 4 \times 10^{16}\ \mathrm{cm}, 6\ M_\odot, 49\ M_\odot)$ and pessimistic scenario $(5\times{10^{52}}\,{\rm{erg}},4 \times 10^{16}\ \mathrm{cm}, 150\,M_\odot, 19\,M_\odot)$. The reconstructed energy for the IC200530 neutrino event  is marked with a dotted vertical. In all  cases, the flux decreases with time with a reduction (growth) of the maximum neutrino energy in the optimistic (pessimistic) scenario.}
	\label{Fig: Temporal fluxes}
\end{figure*}

Figure~\ref{Fig: Temporal fluxes} shows the temporal evolution of the muon neutrino and antineutrino flux for the scenarios with the highest (left panel) and the lowest (right panel) expected number of neutrinos. In all  cases, the flux decreases as time increases and shifts to lower or higher energies, for the most optimistic and pessimistic scenarios, respectively. 
Around the day of detection, the flux in the best scenario is a factor $\mathcal{O}(10^{5})$ larger than the most pessimistic scenario.

\begin{figure}
	\centering
	\includegraphics[width=0.49\textwidth]{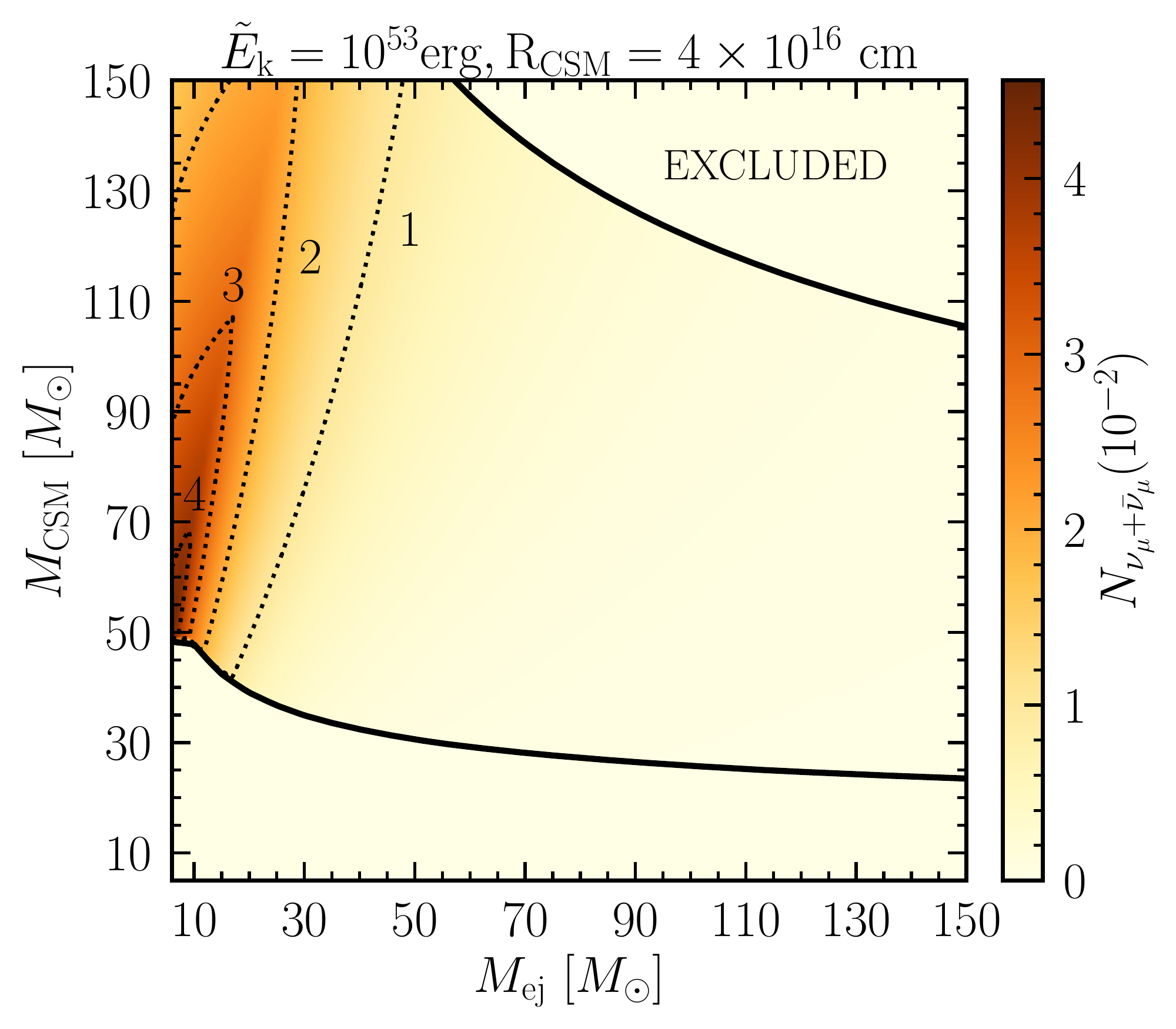}
	\caption{Contour plot of the total number of muon neutrino and antineutrino events expected at the IceCube Neutrino Observatory  from AT2019fdr in $394$~days and in the plane spanned by $ M_{\rm{ej}} $ and $ M_{\rm{CSM}} $  for $\tilde{E}_{\rm{k}}=10^{53}$~erg and ${ R_{\rm{CSM}} = 4 \times 10^{16} }$ cm. The black solid lines marks the allowed region of the parameter space, defined by requiring that the location of the shock at the day of neutrino production is still in the CSM envelope and that the SLSN model parameters are compatible with the the light curve rise time.
	For  fixed $M_{\rm{CSM}}$, the total neutrino number decreases as $M_{\rm{ej}}$ increases, given the decline of the shock velocity $v_{\rm{sh}}$. Viceversa, for fixed $M_{\rm{ej}}$, the number increases as $M_{\rm{CSM}}$ increases, given the larger number of  proton targets for $pp$ interactions. In the region $R_{\rm{bo}}>R_{\rm{dec}}$, one can see an inverted trend. The dotted lines correspond to the contour lines which track the scenarios providing the number of neutrino events displayed, and show how the neutrino number is affected in the transtition from $R_{\rm{bo}}>R_{\rm{dec}}$ to $R_{\rm{bo}}<R_{\rm{dec}}$ regions. See the main text for more details. }
	\label{Fig:contour_plot}
\end{figure}

In order to investigate the origin of IC200530, we integrate the event rate over $394$ days of the neutrino signal for all selected SLSN configurations and obtain the total number of muon neutrino and antineutrino events, $ N_{\nu_{\mu}+\bar{\nu}_{\mu}} $ (Eq.~\ref{eq:Ntot}). A contour plot of $ N_{\nu_{\mu}+\bar{\nu}_{\mu}} $ in the plane spanned by $ M_{\rm{ej}} $ and $ M_{\rm{CSM}} $ is shown in Fig.~\ref{Fig:contour_plot}  for   $ R_{\rm{CSM}}=4\times 10^{16} $~cm and $\tilde{E}_{\rm{k}}=10^{53}$~erg as a representative example.
The allowed region of the parameter space is delimited by the solid black line and plotted in orange (with the color gradient representing a low number of events in lighter hues), while the excluded parameter space is displayed in  light yellow. The dotted contour lines show how the neutrino number is affected as the line $R_{\rm{bo}}=R_{\rm{dec}}$ (along which the cusps of the dotted lines lie) is crossed.

In the region $R_{\rm{dec}}>R_{\rm{bo}}$, for fixed $M_{\rm{ej}}$, the number of neutrino events increases as $M_{\rm{CSM}}$ increases, whilst for fixed $M_{\rm{CSM}}$ and increasing $M_{\rm{ej}}$ we find the opposite trend. The opposite  behaviour occurs for $R_{\rm{dec}}<R_{\rm{bo}}$. 

For the SLSN parameter configurations under consideration, the most optimistic scenarios for the neutrino emission lead to   $ N_{\nu_{\mu} + \bar{\nu}_{\mu}} \simeq 4.6\times 10^{-2}$; the latter is achieved for relatively low values of $ M_{\rm{ej}} =6$--$9 \,M_{\odot}$ and intermediate $ M_{\rm{CSM}} = 49$--$68\, M_{\odot}$, with the best scenario corresponding to $ (\tilde{E}_{\rm{k}}, R_{{\rm{CSM}}},  M_{\rm{ej}}, M_{\rm{CSM}}) \simeq (10^{53}\,{\rm erg}, 4 \times 10^{16}\ {\rm{cm}}, 6\ M_{\odot}, 49\ M_{\odot}) $.

\section{Discussion}
\label{sec:discussion}

\begin{table}[]
    \caption{Number of muon neutrino and antineutrino events expected in $394$~days from the shock breakout from AT2019fdr for the most optimistic and pessimistic scenarios, with the low energy cutoff fixed at $100$~GeV (i.e., excellent discrimination between the astrophysical and atmospheric signals) and $100$~TeV (i.e., under the conservative assumption that the atmospheric background could not be eliminated). The most optimistic and pessimistic scenarios correspond to the following SLSN model parameters: $(\tilde{E}_{\rm{k}},R_{\rm{CSM}}, M_{\rm{ej}},  M_{\rm{CSM}}) = (10^{53}\,{\rm{erg}}, 4 \times 10^{16}\,\mathrm{cm}, 6\,M_\odot, 49\,M_\odot)$ and $(5\times 10^{52}\,{\rm{erg}}, 4 \times 10^{16}\,\mathrm{cm}, 150\,M_\odot, 19\,M_\odot)$, respectively. In the last column we estimate the signalness [$N_{\nu_{\mu}+\bar{\nu}_{\mu}, \rm{astro}}/(N_{\nu_{\mu}+\bar{\nu}_{\mu}, \rm{astro}}+N_{\nu_{\mu}+\bar{\nu}_{\mu}, \rm{atm}})$], by computing  the number of atmospheric neutrino events over a period of $394$~days, for $0.75^{\circ}$ around the direction of the source.  \label{table:range_of_nu_events}}
    \begin{center}
    \hspace{-0.9cm}
    \begin{tabular}[c]{ccc|c}
        \toprule
        Energy cut & $N_{\nu_{\mu}+\bar{\nu}_{\mu},\ \mathrm{pessimistic}}$ & $N_{\nu_{\mu}+\bar{\nu}_{\mu},\ \mathrm{optimistic}}$& Signalness \\
        \hline
        $E_{\nu,\rm{min}}= 100$~GeV &  $8\times 10^{-6}$& $ 4.6\times 10^{-2}$ & $10^{-4}$--$1\%$  \\
        $E_{\nu,\rm{min}}= 100$~TeV &  $ 9.5\times 10^{-9}$& $ 4.6\times 10^{-3}$ & $10^{-4}$--$\, 40\%$ \\
        \toprule
	\end{tabular}
	\end{center}
\end{table}

Table~\ref{table:range_of_nu_events} summarizes the total number of muon neutrino and antineutrino events expected within $394$~days from the shock breakout from AT2019fdr for the most optimistic and pessimistic SLSN configurations in terms of neutrino emission. The largest [smallest] number of events is obtained for the SLSN configuration with
$(\tilde{E}_{\rm{k}},R_{\rm{CSM}}, M_{\rm{ej}},  M_{\rm{CSM}}) = (10^{53}~\mathrm{erg}, 4 \times 10^{16}\,\mathrm{cm}, 6\,M_\odot, 49\,M_\odot)$ [$(5\times 10^{52}\,{\rm{erg}}, 4 \times 10^{16}\,\mathrm{cm}, 150\,M_\odot, 19\,M_\odot)$],
and correspond to the edges of the band in Fig.~\ref{Fig:N_numu_rate_band}.

An important aspect to consider in the interpretation of the neutrino event IC200530 concerns the discrimination of the atmospheric neutrino background, which dominates over the astrophysical neutrino flux below $\simeq 100$~TeV. As such, in Table~\ref{table:range_of_nu_events} we distinguish   between one  case with  the lower energy cutoff fixed at $100$~GeV, mimicking  excellent discrimination of the atmospheric neutrino background (see Sec.~\ref{sec:nurateflux}), and one more conservative case  with the lower energy cutoff at $100$~TeV. The latter case reproduces a situation where  the atmospheric neutrino events could not be distinguished from the astrophysical ones in the lower energy range. Although a dedicate likelyhood analysis is beyond the scope of this work,  the last column of Table~\ref{table:range_of_nu_events}  reports $N_{\nu_{\mu}+\bar{\nu}_{\mu}, \rm{astro}}/(N_{\nu_{\mu}+\bar{\nu}_{\mu}, \rm{astro}}+N_{\nu_{\mu}+\bar{\nu}_{\mu}, \rm{atm}})$, which should give an idea of  the expected signalness and gives an indication of the  probability that a detected neutrino event could be of  astrophysical origin. It is evident that only an optimal discrimination of the atmospheric neutrino background allows to obtain  a signalness of  $40\%$, roughly comparable with the one of the neutrino event IC200530.
The evolution of the neutrino curve shown in Fig.~\ref{Fig:N_numu_rate_band} should  be  considered carefully. In fact, some intermediate SLSN scenarios enclosed in the envelope in  Fig.~\ref{Fig:N_numu_rate_band}, and compatible with the  reconstructed energy of the neutrino event IC200530A, have an event rate still increasing at the day of detection, therefore increasing the neutrino detection chances at later times, as it is the case for the neutrino event IC200530.

In order to assess whether the number of expected events in Table~\ref{table:range_of_nu_events} is compatible with the detection of one neutrino event (IC200530) after $394$~days from the shock breakout, we take into account the Eddington bias on neutrino observations. 
The Eddington bias must be taken into account when dealing with very small number of neutrino events, such as in this case; we refer the interest reader to \cite{Strotjohann:2018ufz} for a dedicated discussion.
By relying on the local rate of  SLSN IIn provided in~\cite{Quimby:2013jb}  and integrating over the cosmic history by assuming that the redshift evolution of SLSN IIn follows the star formation rate~\citep{Yuksel:2008cu}, we obtain an average effective density of SLSN IIn equal to $\mathcal{O}(3 \times 10^3)$~Mpc$^{-3}$. Although Fig.~2 of \citet{Strotjohann:2018ufz} was derived within a simplified framework and for constant redshift evolution, by extrapolating to larger effective source densities we conclude that  the number of expected events   in Table~\ref{table:range_of_nu_events} may be compatible with the detection of at least one or two neutrino events from AT2019fdr. By taking into account the fact that the neutrino energy distribution of AT2019fdr falls in a region where the discrimination of the atmospheric neutrino background may be challenging, our findings hint towards a possible association of  the neutrino event IC200530  to AT2019fdr. 
In addition, our results are compatible with the upper limits on the neutrino emission from the AT2019fdr source  placed by the ANTARES Collaboration~\citep{ANTARES:2021jmp}.

We should stress that the forecasted number of expected neutrino events includes some caveats related to our modeling. For example, 
one of the sources of uncertainty in the computation of the neutrino flux is  the proton acceleration efficiency $\varepsilon_{p}$. In this paper, we have adopted an optimistic $\varepsilon_{p}=0.1$, assuming that  the shocks accelerating protons are parallel or quasi-parallel and therefore efficient diffusive shock acceleration occurs~\citep{Caprioli:2013dca}. However, lower values of $\varepsilon_{p}$ would be possible for oblique shocks, with poorer particle acceleration efficiency.  
Values as low as $\varepsilon_{p}\simeq 0.003$--$0.01$ have been inferred in \cite{Aydi:2020znu} for a  nova,  assuming shocks as the powering source of the simultaneously observed optical and $\gamma$-rays. However,  observational constraints from other optical transients, including SLSNe, are still lacking; in addition,  AT2019fdr is much more luminous than classical novae, possibly hinting to different conditions present in the acceleration region. 

We  stress that the neutrino flux scales linearly with $\varepsilon_{p}$, allowing the reader to easily scale our results. All cases summarized in  Table~\ref{table:range_of_nu_events} would  be compatible with the detection of one neutrino event, after taking into account the Eddington bias. Indeed, the detection of a single neutrino event may  actually hint towards  intermediate SLSN configurations, as well as  values of $\varepsilon_{p}$ lower than our benchmark one.

Similarly, in this work we have assumed that protons are accelerated at the shock to a power law with slope $k=2$, which is predicted by the test particle theory of diffusive shock acceleration. Nonetheless, non-linear effects involving the amplified magnetic field can kick in, modifying the shock structure and making the cosmic ray spectra mildly steeper than $k=2$~\citep{Caprioli:2021nwv}. Larger $k$ would result in steeper neutrino spectra, since the latter inherit the shape of the parent proton spectrum; as a consequence,  lower fluxes should be expected in  the energy of interest.

Another caveat to take into account concerns the use of the AT2019fdr discovery date in the observer frame as the breakout time of the shock. In fact, based on the non-detections in the ZTF data, we have assumed an explosion epoch at the first detection at MJD$\, = 58606 \pm 6$~days on the basis of a fit on the existing data.  Nevertheless, even allowing for an onset of the shock breakout to be as much as $\sim 20$ days earlier than the first observed light, our predictions in Table~\ref{table:range_of_nu_events} would not be affected by a factor larger than $10\%$.

Since initial submission of this manuscript, other publications have analysed IC200530 under the paradigm of a TDE origin \citep{Reusch:2021ztx}. The additional data presented within these works suggest that an apparent increase in the late time near infrared (NIR) emission may be attributed to a dust from the TDE emission. However, increasing late time NIR emission has been seen in other interacting SNe. For instance, the bright SN IIn SN2010jl exhibits such a NIR increase at late times;  high-resolution spectroscopic observations show that this increasing emission was the result of rapid dust formation within the SN ejecta~\citep{Gall2014}.

In addition, the vast majority of TDEs show bright X-ray emission  over the full optical/UV evolution of an event \citep[e.g.,][]{2017ApJ...838..149A, 2017MNRAS.466.4904B, 2021MNRAS.500.1673H, 2021ApJ...912..151W}. Of those whose emission is dominated by optical/UV but has been detected in X-rays, the X-ray luminosities are an order of magnitude or more fainter than the eROSITA detection \citep[e.g.,][]{2020ApJ...889..166J,2019ApJ...883..111H, 2020ApJ...903...31H, 2021ApJ...917....9H}. In addition, AT2019fdr is found close to the nucleus in a Narrow-Line Seyfert 1 active galaxy \citep{2020arXiv201008554F}. Seyfert AGN galaxies are known to exhibit bright X-rays, with a mean X-ray luminosity of $\sim10^{43}$ erg s$^{-1}$ \citep[e.g.,][]{2017ApJS..233...17R} similar to that detected by eROSITA. Furthermore, \cite{2017ApJS..233...17R} and references therewithin showed that a significant fraction of un-obscured AGN, and particularly those selected in optical, tend to exhibit excess soft X-ray emission that can be best described by an absorbed blackbody. They found that this excess can be well fit with a rest-frame blackbody temperature ranging between $\sim0.5$--$0.25$~keV, with a mean temperature of $\sim 0.1$~keV, which is consistent with the blackbody temperature derived by \cite{Reusch:2021ztx}. Due to the angular resolution of eROSITA, further high resolution X-ray observations would be necessary to confirm whether the detected X-ray emission  arises from its host galaxy's AGN or is consistent with the location of AT2019fdr.

If the latter was the case, a detection of X-rays from a SLSN at late times would not be surprising. The total luminosity of the shock and the pre-shock column density of the CSM medium  determine the observation features of high-energy radiation. Unless we are in the presence of  extremely high shock temperatures or a high ratio of the shock luminosity to the column density, which would guarantee the CSM ionization to a large extent, the photoelectric absorption is an important energy loss channel for high energy photons. The latter could explain the non observation of X-rays at earlier times~\citep{Pan:2013nfa}. Unfortunately, as  already discussed, there could  be degeneracies  among the parameters, leading to similar properties of the SLSN light curve. Nevertheless, the slow rise of the optical light curve, the very high luminosity peak, and the non observation of X-rays for a considerable amount of time after the first detection would  point towards scenarios with highly energetic and relatively low mass ejecta moving through extended high CSM mass stellar winds or shells.

\section{Conclusions}
\label{sec:conclusions}

The  IceCube neutrino event IC200530 has been proposed to be in likely coincidence with the source AT2019fdr located at $z=0.2666$, observed in the ultraviolet and optical bands, and  interpreted as a tidal distruption event candidate in a Narrow-Line Seyfert 1 galaxy. In this paper, we show that  the spectra, light curve and color evolution of   AT2019fdr may be compatible with the ones of a hydrogen rich superluminous supernova  instead. 

Under this assumption, the neutrino event  IC200530, detected $\sim 300$~days after the peak of the electromagnetic emission and with a reconstructed energy of $80$~TeV,  may have originated as a result of inelastic proton-proton collisions due to the interaction of the supernova ejecta with the circumstellar medium. 
We find that approximately $10^{-8}$--$5 \times 10^{-2}$ muon neutrino and antineutrino events could have been produced by AT2019fdr within the timeframe of interest ( see Table~\ref{table:range_of_nu_events}), by taking into account the uncertainties on the total ejecta energetics, ejecta mass and on the  properties of the the circumstellar medium, as well as the uncertainties in the discrimination of the atmospheric and astrophysical neutrino fluxes.  By considering the Eddington bias on neutrino observations, our findings may be compatible with the detection of one neutrino event from   AT2019fdr.

In conclusion, the neutrino event IC200530 may be associated with the hydrogen rich superluminous supernova AT2019fdr.  As a deeper understanding of the electromagnetic data will become available, neutrinos could be powerful messengers to help to disentangle the nature of AT2019fdr and discover its physics.

\acknowledgments
We thank Markus Ahlers and Anna Franckowiak for insightful discussions. This project has received funding from the  Villum Foundation (Projects No.s~13164, 37358, 16599, and 25501), the Carlsberg Foundation (CF18-0183), the Deutsche Forschungsgemeinschaft through Sonderforschungbereich
SFB~1258 ``Neutrinos and Dark Matter in Astro- and Particle Physics'' (NDM), and the Australian Research Council Centre of Excellence for All Sky Astrophysics in 3 Dimensions (ASTRO 3D), through project number CE170100013. 
We acknowledge the use of public data from the \textit{Swift} data archive. The results presented in this paper are based on observations obtained with the Samuel Oschin 48-inch Telescope at the Palomar Observatory as part of the Zwicky Transient Facility project. The Zwicky Transient Facility is supported by the National Science Foundation under Grant No.~AST-1440341 and a collaboration including Caltech, the Infrared Processing and Analysis Center (IPAC), the Weizmann Institute for Science, the Oskar Klein Center at Stockholm University, the University of Maryland, the University of Washington, Deutsches Elektronen-Synchrotron and Humboldt University, Los Alamos National Laboratories, the TANGO Consortium of Taiwan, the University of Wisconsin (UW) at Milwaukee, and Lawrence Berkeley National Laboratories. Operations are conducted by Caltech Optical Observatories, IPAC, and UW. This work has also made use of data from the Asteroid Terrestrial-impact Last Alert System (ATLAS) project. The ATLAS project is primarily funded to search for near Earth asteroids through NASA grants NN12AR55G, 80NSSC18K0284, and 80NSSC18K1575; byproducts of the Near Earth Asteroid (NEO) search include images and catalogs from the survey area. This work was partially funded by Kepler/K2 grant J1944/80NSSC19K0112 and HST GO-15889, and STFC grants ST/T000198/1 and ST/S006109/1. The ATLAS science products have been made possible through the contributions of the University of Hawaii Institute for Astronomy, the Queen’s University Belfast, the Space Telescope Science Institute, the South African Astronomical Observatory, and the Millennium Institute of Astrophysics, Chile.

\appendix
\section{Parameter space adopted in the modeling of AT2019\small{fdr}}
\label{Appendix: parameter space}
In this Appendix we investigate how the space of the AT2019fdr parameters reported in Table~\ref{table:parameters} is constrained by our two selection criteria: 1)  the time necessary for the forward shock to cross the CSM envelope between $R_{\rm{bo}}$ and $R_{\rm{CSM}}$ is at least $394$~days, and 2)  the rising time to the peak of the bolometric lightcurve (see Fig.~\ref{Fig:LC_SN}) is $98$~days in the observer frame. 
\begin{figure*}
	\begin{center}
		\includegraphics[width=0.45\textwidth]{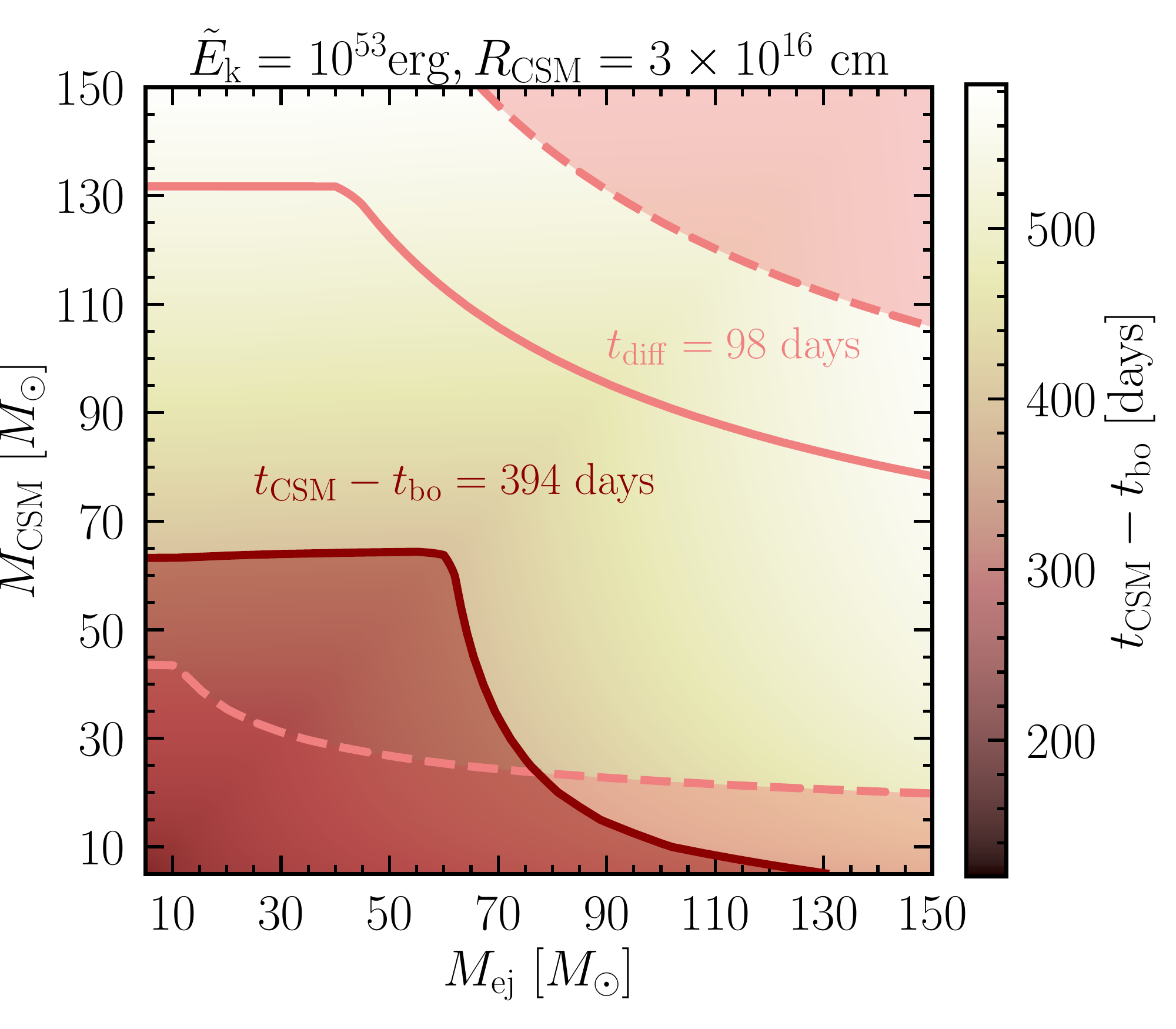}
		\includegraphics[width=0.45\textwidth]{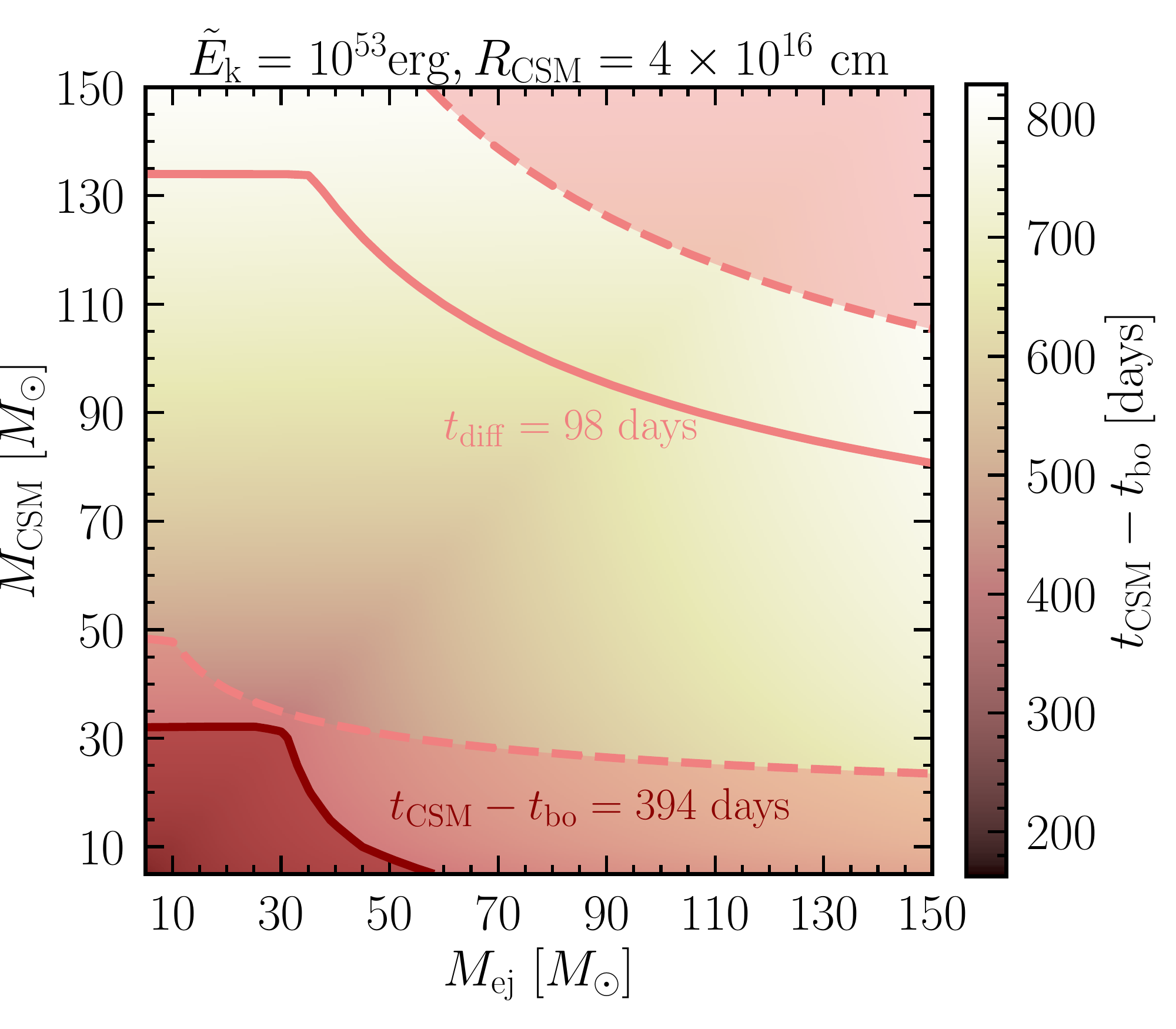}
	\end{center}
	\caption{\textit{Left panel}: Contour plot of the time the shock takes  to travel from $R_{\rm{bo}}$ to $R_{\rm{CSM}}$ in the plane spanned by $M_{\rm{rj}}$ and $M_{\rm{CSM}}$. The solid bordeaux line  constrains the allowed  parameter space by requiring that $t_{\rm{CSM}}-t_{\rm{bo}} \ge 394$~days (solid bordeaux line). 
	The dashed pink lines constrain the allowed parameter space by requiring that  the rising time of the AT2019fdr lightcurve is compatible within a $50\%$ uncertainty  with the analytic estimate of the diffusion time provided in  Eq.~\ref{Eq: t_diff}; the latter is represented by the solid pink line. {\textit {Right panel}:} The same as in the left panel, but for $R_{\rm{CSM}}=4\times 10^{16}$~cm. For larger $R_{\rm{CSM}}$, the   crossing time constraint becomes looser, whilst the one related to $t_{\rm{diff}}$ slowly becomes more stringent.}
	\label{Fig: t_csm-t_bo}
\end{figure*}

Because of the approximations  involved in the definition of $t_{\rm{diff}}$ in Eq.~\ref{Eq: t_diff}, we take into account an uncertainty of $50\%$ on the diffusion time. Figure~\ref{Fig: t_csm-t_bo} shows a contour plot of the time that the shock takes  to travel from $R_{\rm{bo}}$ to $R_{\rm{CSM}}$ for $\tilde{E}_{\rm{k}}=10^{53}$~erg. We can see that
the smaller the CSM width, the shorter the time it takes for the shock to reach  $R_{\rm{CSM}}$. Indeed, in the left panel of Fig.~\ref{Fig: t_csm-t_bo}, as opposed to the right one, almost half of the SLSN configurations with $M_{\rm{ej}}\lesssim 70\,M_{\odot}$ and $M_{\rm{CSM}}\lesssim 70\,M_{\odot}$ are excluded. This is mainly due to the fact that  $R_{\rm{bo}}\ll R_{\rm{dec}}$ for   $M_{\rm{ej}}/M_{\rm{CSM}}$ that is not large, implying that most of the evolution of the shock in the CSM is in the free expansion phase (see Eq.~\ref{Eq: Rsh in free phase}), thus with larger velocities. Furthermore, this criterion completely excludes all the configurations with $R_{\rm{CSM}}=2\times 10^{16}$~cm and $\tilde{E}_{\rm{k}}=10^{53}$~erg. As $R_{\rm{CSM}}$ increases (see the right panel of Fig.~\ref{Fig: t_csm-t_bo}), the most stringent constraint comes from the compatibility of $t_{\rm{diff}}$ with the observed light curve. 

The same trend holds for the case with $\tilde{E}_{\rm{k}}=5\times 10^{52}$~erg (not shown here), with the difference that  there are compatible scenarios with our requirements already for $R_{\rm{CSM}}=2\times 10^{16}$~cm. For this latter case, for fixed $M_{\rm{ej}}, M_{\rm{CSM}}$ and $R_{\rm{CSM}}$, the shock velocity $v_{\rm{sh}}$ is lower, allowing for longer times required to  cross the CSM.

\section{Maximum proton energy}
\label{Appendix: maximum_proton_energy}
\begin{figure*}
	\begin{center}
		\includegraphics[width=0.45\textwidth]{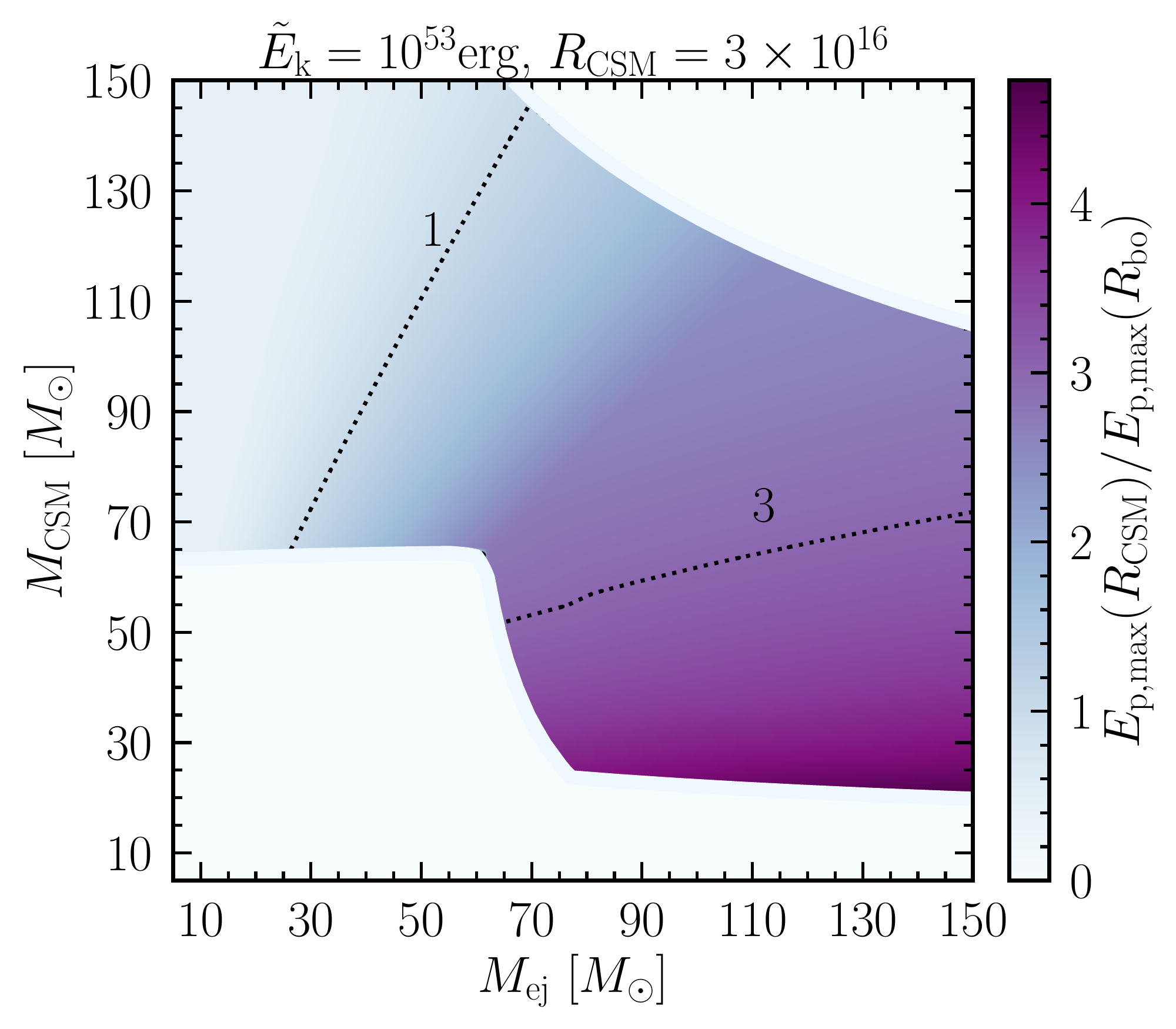}
		\includegraphics[width=0.45\textwidth]{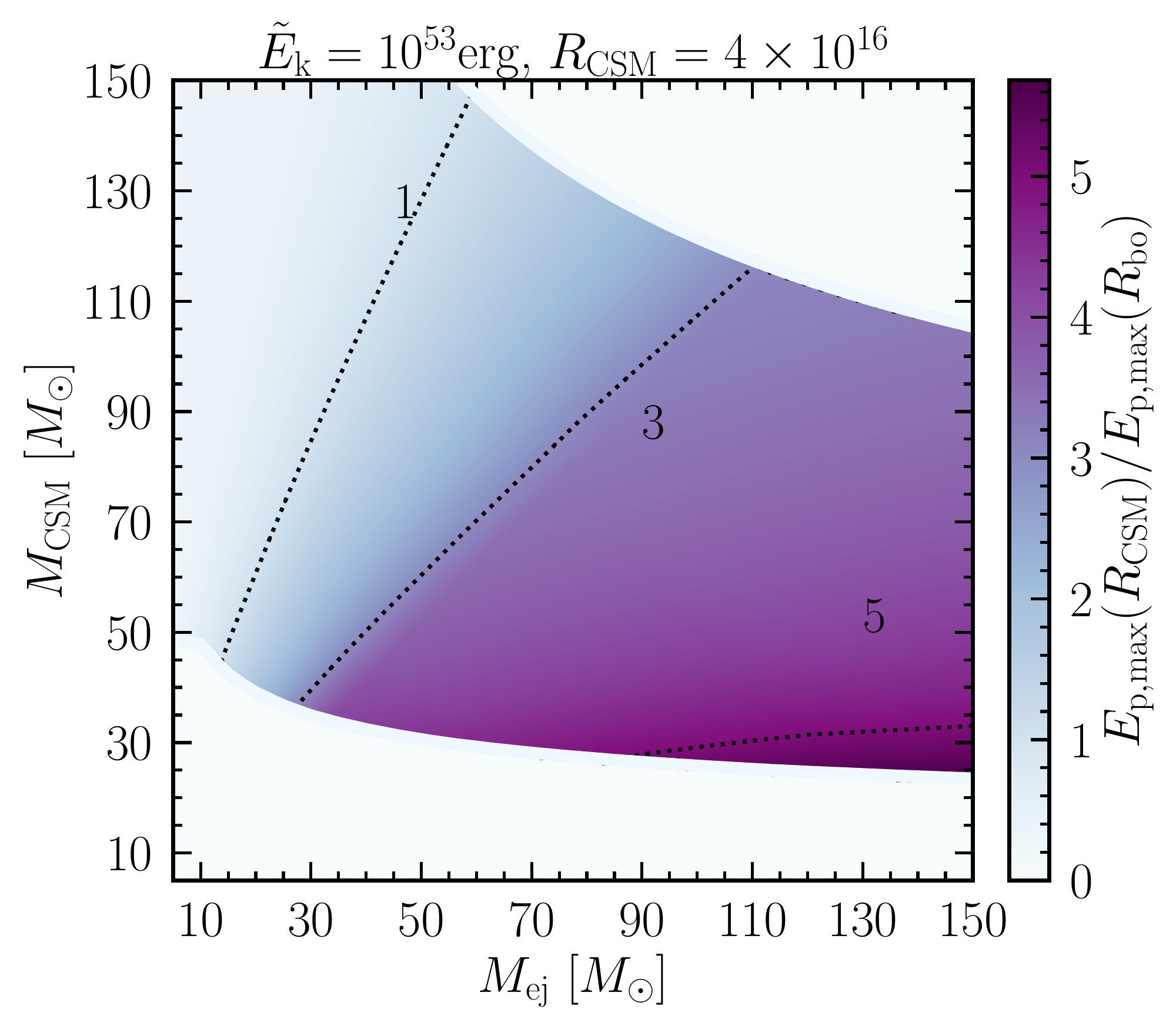}
	\end{center}
	\caption{\textit{Left panel}: Contour plot of the ratio between the  maximum proton energy $E_{\rm{p,max}}$ at $R_{\rm{CSM}}=3\times10^{16}$~cm and at the breakout radius $R_{\rm{bo}}$ in  the plane spanned by $M_{\rm{ej}}$ and $M_{\rm{CSM}}$. 
	For relatively low values of $M_{\rm{ej}}$ with respect to $M_{\rm{CSM}}$, this ratio tends to decrease. This is due to the fact that for very low $M_{\rm{ej}}/M_{\rm{CSM}}$, $R_{\rm{dec}}< R_{\rm{bo}}$, causing a  fast drop of $E_{\rm{p,max}}$. Viceversa, for very large $M_{\rm{ej}}/M_{\rm{CSM}}$, the deceleration always occurs  at $R>R_{\rm{CSM}}$, allowing for a continual increase of $E_{\rm{p,max}}$ as the time goes by.  Intermediate values of $M_{\rm{ej}}/M_{\rm{CSM}}$ lead to intermediate trends, with the free expansion and decelerating phase both being present between $R_{\rm{bo}}$ and $R_{\rm{CSM}}$. The dotted black lines indicate the regions the ratio is larger than $1$ and  $3$. \textit{Right panel:} The same as in the left panel, but for a larger $R_{\rm{CSM}}$. The effect of increasing $R_{\rm{CSM}}$, while keeping fixed all the other parameters, is to decrease the CSM density and thus to allow for larger $E_{\rm{p,max}}$, since the $pp$ interactions become less efficient. }
	\label{Fig: Ep_max_ratio}
\end{figure*}

In this appendix, we explore the temporal evolution of $E_{\rm{p,max}}$ for the set of parameters $\tilde{E}_{\rm{k}}, R_{\rm{CSM}}, M_{\rm{ej}}$ and $M_{\rm{CSM}}$ considered in this work (see Table~\ref{table:parameters}). We provide an idea of the behaviour of $E_{\rm{p,max}}$ by displaying in Fig.~\ref{Fig: Ep_max_ratio} the ratio between its value at the CSM radius $R_{\rm{CSM}}$ and the breakout radius $R_{\rm{bo}}$, for  $\tilde{E}_{\rm{k}}=10^{53}$~erg with $R_{\rm{CSM}}=3\times 10^{16}$~cm (left panel) and $R_{\rm{CSM}}=4 \times 10^{16}$~cm (right panel). In both cases, the region where $E_{\rm{p,max}}(R_{\rm{CSM}})/E_{\rm{p,max}}(R_{\rm{bo}})<1$ is the one with relatively low values of $M_{\rm{ej}}/M_{\rm{CSM}}$. Here, either $R_{\rm{bo}}>R_{\rm{dec}}$ or $R_{\rm{bo}}\lesssim R_{\rm{dec}}$, meaning that most of the shock evolution occurs in the decelerating phase (see Eq.~\ref{Eq: R_sh in dec phase}). When this is the case, the acceleration efficiency drops at a faster rate, leading to decreasing $E_{\rm{p,max}}$ (see Eq.\ref{eq:Ep_max}).

On the other hand, for  large  $M_{\rm{ej}}/M_{\rm{CSM}}$, $R_{\rm{dec}}>R_{\rm{CSM}}$ is satisfied, implying an increase of $E_{\rm{p,max}}$. The intermediate regimes [$M_{\rm{ej}}/M_{\rm{CSM}}\sim \mathcal{O}(1)$] are those in which both free expansion and deceleration occur between $R_{\rm{bo}}$ and $R_{\rm{CSM}}$, being the latter shorter compared to the former, and thus leaving the  tendency of $E_{\rm{p,max}}(R_{\rm{CSM}})/E_{\rm{p,max}}(R_{\rm{bo}})$ to increase  unaffected. By keeping  $\tilde{E}_{\rm{k}}, M_{\rm{ej}}$ and $M_{\rm{CSM}}$ fixed, a larger $R_{\rm{CSM}}$ leads to a lower CSM density, with longer $t_{\rm{pp}}$; thus, a larger $E_{\rm{p,max}}(R_{\rm{CSM}})$ is achievable. This effect is more significant than the slight increase of $E_{\rm{p,max}}(R_{\rm{bo}})$ for larger $R_{\rm{CSM}}$.

Finally, lower values of $\tilde{E}_{k}$ do not compromise the overall trend outlined above. The only effect of decreasing the energy, whilst keeping  all other parameters fixed, is to  reduce $v_{\rm{sh}}$ (see Eq.~\ref{Eq: Rsh in free phase}) and in turn the acceleration rate, which result in overall smaller values of $E_{\rm{p,max}}$.

\newpage
\bibliography{references_SLSN}

\begin{thebibliography}{}
\expandafter\ifx\csname natexlab\endcsname\relax\def\natexlab#1{#1}\fi
\providecommand{\url}[1]{\href{#1}{#1}}
\providecommand{\dodoi}[1]{doi:~\href{http://doi.org/#1}{\nolinkurl{#1}}}
\providecommand{\doeprint}[1]{\href{http://ascl.net/#1}{\nolinkurl{http://ascl.net/#1}}}
\providecommand{\doarXiv}[1]{\href{https://arxiv.org/abs/#1}{\nolinkurl{https://arxiv.org/abs/#1}}}

\bibitem[{Aartsen {et~al.}(2017)}]{IceCube:2016qvd}
Aartsen, M.~G., {et~al.} 2017, Astrophys. J., 835, 45,
  \dodoi{10.3847/1538-4357/835/1/45}

\bibitem[{Aartsen {et~al.}(2018{\natexlab{a}})}]{IceCube:2018fhm}
---. 2018{\natexlab{a}}, Phys. Rev. D, 98, 062003,
  \dodoi{10.1103/PhysRevD.98.062003}

\bibitem[{Aartsen {et~al.}(2018{\natexlab{b}})}]{IceCube:2018dnn}
---. 2018{\natexlab{b}}, Science, 361, eaat1378,
  \dodoi{10.1126/science.aat1378}

\bibitem[{Aartsen {et~al.}(2020)}]{IceCube:2020acn}
---. 2020, Phys. Rev. Lett., 125, 121104,
  \dodoi{10.1103/PhysRevLett.125.121104}

\bibitem[{Abbasi {et~al.}(2021{\natexlab{a}})}]{IceCube:2021xar}
Abbasi, R., {et~al.} 2021{\natexlab{a}}, \dodoi{10.21234/CPKQ-K003}

\bibitem[{Abbasi {et~al.}(2021{\natexlab{b}})}]{IceCube:2020wum}
---. 2021{\natexlab{b}}, Phys. Rev. D, 104, 022002,
  \dodoi{10.1103/PhysRevD.104.022002}

\bibitem[{Abbasi {et~al.}(2021{\natexlab{c}})}]{IceCube:2020mzw}
---. 2021{\natexlab{c}}, Astrophys. J., 910, 4,
  \dodoi{10.3847/1538-4357/abe123}

\bibitem[{Acciari {et~al.}(2021)Acciari, Ansoldi, Antonelli, Arbet~Engels,
  Artero, Asano, Baack, Babic, Baquero, Barres~de Almeida, Barrio, Batkovi?,
  Becerra~Gonzalez, Bednarek, Bellizzi, Bernardini, Bernardos, Berti,
  Besenrieder, Bhattacharyya, Bigongiari, Biland, Blanch, Bökenkamp, Bonnoli,
  Bosnjak, Busetto, Carosi, Ceribella, Cerruti, Chai, Chilingarian, Cikota,
  Colak, Colombo, Contreras, Cortina, Covino, D'Amico, D'Elia, Da~Vela, Dazzi,
  De~Angelis, De~Lotto, Delfino, Delgado, Delgado~Mendez, Depaoli, Di~Pierro,
  Di~Venere, Do~Souto~Espiñeira, Dominis~Prester, Donini, Dorner, Doro,
  Elsaesser, Fallah~Ramazani, Fattorini, Fonseca, Font, Fruck, Fukami,
  Fukazawa, García~López, Garczarczyk, Gasparyan, Gaug, Giglietto, Giordano,
  Gliwny, Godinovic, Green, Green, Hadasch, Hahn, Heckmann, Herrera, Hoang,
  Hrupec, Hütten, Inada, Ishio, Iwamura, Jiménez~Martínez, Jormanainen,
  Jouvin, Karjalainen, Kerszberg, Kobayashi, Kubo, Kushida, Lamastra, LELAS,
  Leone, Lindfors, Linhoff, Lombardi, Longo, Lopez-Coto, López-Moya,
  López-Oramas, Loporchio, Machado~de Oliveira~Fraga, Maggio, Majumdar,
  MAKARIEV, Mallamaci, Maneva, Manganaro, Mannheim, Maraschi, Mariotti,
  Martinez, Mazin, Menchiari, Mender, Mi?anovi?, Miceli, Miener, Miranda,
  Mirzoyan, Molina, Moralejo, Morcuende, Moreno, Moretti, Nakamori, Nava,
  Neustroev, Nigro, Nilsson, Nishijima, Noda, Nozaki, Ohtani, Oka,
  Otero-Santos, Paiano, Palatiello, Paneque, Paoletti, Paredes, Pavleti?,
  Peñil, Persic, Pihet, Prada~Moroni, Prandini, Priyadarshi, Puljak, Rhode,
  Ribó, Rico, Righi, Rugliancich, Sahakyan, Saito, Sakurai, Satalecka,
  Saturni, Schleicher, Schmidt, Schweizer, Sitarek, ¦nidari?, Sobczy?ska,
  Spolon, Stamerra, Stri¨kovi?, Strom, Strzys, Suda, Suri?, Takahashi,
  Takeishi, Tavecchio, Temnikov, Terzic, Teshima, Tosti, Truzzi, Tutone, Ubach,
  van Scherpenberg, Vanzo, VAZQUEZ~ACOSTA, Ventura, VERGUILOV, Vigorito,
  Vitale, Vovk, Will, Wunderlich, Yamamoto, Zari?, Balbo, Bretz, Buss,
  Eisenberger, Hildebrand, Iotov, Kalenski, Neise, Noethe, Paravac, Sliusar,
  Walter, Abbasi, Ackermann, Adams, Aguilar, Ahlers, Ahrens, Alispach,
  Alves~Junior, Amin, An, Andeen, Anderson, Anton, Arguelles, Ashida, Axani,
  Bai, Balagopal~V., Barbano, Barwick, Bastian, Basu, Baur, Bay, Beatty,
  Becker, Becker~Tjus, Bellenghi, BenZvi, Berley, Besson, Binder, Bindig,
  Blaufuss, Blot, Boddenberg, Bontempo, Borowka, Boser, Botner, Bottcher,
  Bourbeau, Bradascio, Braun, Bron, Brostean-Kaiser, Browne, Burgman, Burley,
  Busse, Campana, Carnie-Bronca, Chen, Chirkin, Choi, Clark, Clark, Classen,
  Coleman, Collin, Conrad, Coppin, Correa, Cowen, Cross, Dappen, Dave,
  DE~CLERCQ, DeLaunay, Dembinski, Deoskar, De~Ridder, Desai, Desiati, de~Vries,
  de~Wasseige, De~With, DeYoung, Dharani, Diaz, Diaz-Velez, Dittmer, Dujmovic,
  Dunkman, DuVernois, Dvorak, Ehrhardt, Eller, Engel, Erpenbeck, Evans,
  Evenson, Fan, Fazely, Fiedlschuster, Fienberg, Filimonov, Finley, Fischer,
  Fox, Franckowiak, Friedman, Fritz, Furst, Gaisser, Gallagher, Ganster,
  Garcia, Garrappa, Gerhardt, Ghadimi, Glaser, Glauch, Glusenkamp, Goldschmidt,
  Gonzalez, Goswami, Grant, Grégoire, Griswold, Gunduz, Günther, Haack,
  Hallgren, Halliday, Halve, Halzen, Ha~Minh, Hanson, Hardin, Harnisch, Haungs,
  Hauser, Hebecker, Helbing, Henningsen, Hettinger, Hickford, Hignight, Hill,
  Hill, Hoffman, Hoffmann, Hoinka, Hokanson-Fasig, Hoshina, Huang, Huber,
  Huber, Hultqvist, Hunnefeld, Hussain, In, Iovine, Ishihara, Jansson,
  Japaridze, Jeong, Jones, Kang, Kang, Kang, Kappes, Kappesser, Karg, Karl,
  Karle, Katz, Kauer, Kellermann, Kelley, Kheirandish, Kin, Kintscher, Kiryluk,
  Klein, Koirala, Kolanoski, Kontrimas, Kopke, Kopper, Kopper, Koskinen,
  Koundal, Kovacevich, Kowalski, Kozynets, Kun, Kurahashi, Lad, Lagunas~Gualda,
  Lanfranchi, Larson, Lauber, Lazar, Lee, Leonard, Leszczy?ska, Li, Lincetto,
  Liu, Liubarska, Lohfink, Lozano~Mariscal, Lu, Lucarelli, Ludwig, Luszczak,
  Lyu, Ma, Madsen, Mahn, Makino, Mancina, Maris, Maruyama, Mase, McElroy,
  McNally, Mead, Meagher, Medina, Meier, Meighen-Berger, Micallef, Mockler,
  Montaruli, Moore, Morse, Moulai, Naab, Nagai, Naumann, Necker, Nguyen,
  Niederhausen, Nisa, Nowicki, Nygren, Obertacke~Pollmann, Oehler, Olivas,
  O'Sullivan, Pandya, Pankova, Park, Parker, Paudel, Paul, Perez de~los Heros,
  Peters, Philippen, Pieloth, Pieper, Pittermann, Pizzuto, Plum, Popovych,
  Porcelli, Prado~Rodriguez, Price, Pries, Przybylski, Raab, Raissi, Rameez,
  Rawlins, Rea, Rehman, Reimann, Renzi, Resconi, Reusch, Richman, Riedel,
  Roberts, Robertson, Roellinghoff, Rongen, Rott, Ruhe, Ryckbosch,
  Rysewyk~Cantu, Safa, Saffer, Sanchez~Herrera, Sandrock, Sandroos, Santander,
  Sarkar, Sarkar, Scharf, Schaufel, Schieler, Schindler, Schlunder, Schmidt,
  Schneider, Schneider, Schröder, Schumacher, Schwefer, Sclafani, Seckel,
  Seunarine, Sharma, Shefali, Silva, Skrzypek, Smithers, Snihur, Soedingrekso,
  Soldin, Spannfellner, Spiczak, Spiering, Stachurska, Stamatikos, Stanev,
  Stein, Stettner, Steuer, Stezelberger, Sturwald, Stuttard, Sullivan, Taboada,
  Tenholt, Ter-Antonyan, Tilav, Tischbein, Tollefson, Tönnis, Toscano, Tosi,
  Trettin, Tselengidou, Tung, Turcati, Turcotte, Turley, Twagirayezu, Ty,
  Unland~Elorrieta, Valtonen-Mattila, Vandenbroucke, van Eijndhoven, Vannerom,
  van Santen, Verpoest, Vraeghe, Walck, Watson, Weaver, Weigel, Weindl, Weiss,
  Weldert, Wendt, Werthebach, Weyrauch, Whitehorn, Wiebusch, Williams, Wolf,
  Woschnagg, Wrede, Wulff, Xu, Xu, Yanez, Yoshida, Yu, Yuan, Zhang, Jin,
  Abdalla, Aharonian, Ait-Benkhali, Anguener, Arcaro, Armand, Armstrong,
  Ashkar, Backes, Baghmanyan, Barbosa~Martins, Barnacka, Barnard, Batzofin,
  Becherini, Berge, Bernlöhr, Bi, Boettcher, Boisson, Bolmont, de~Bony,
  Breuhaus, Brose, Brun, Bulik, Bylund, Cangemi, Caroff, Casanova, Catalano,
  Chambery, Chand, Chen, Cotter, Curlo, Damascene~Mbarubucyeye, Davids, Davies,
  Devin, Djannati-Ataï, Dmytriiev, Donath, Doroshenko, Dreyer, Du~Plessis,
  Duffy, Egberts, Einecke, Emery, ERNENWEIN, Fegan, Feijen, Fiasson, Fichet~de
  Clairfontaine, Fontaine, Frans, Fuessling, Funk, Gabici, Gallant,
  Ghafourizade, Giavitto, Giunti, Glawion, Glicenstein, Grondin, Hattingh,
  Haupt, HERMANN, Hinton, Hofmann, Hoischen, Holch, Holler, Horns, Huang,
  Huber, Hörbe, Jamrozy, Jankowsky, Joshi, JUNG, Kasai, Katarzynski, Katz,
  Khangulyan, Khelifi, Klepser, Kluzniak, Komin, Konno, Kosack, Kostunin,
  Kreter, Kukec~Mezek, Kundu, Lamanna, Le~Stum, Lemiere, Lemoine-Goumard,
  Lenain, Leuschner, Levy, Lohse, Luashvili, Lypova, Mackey, Majumdar,
  Malyshev, MALYSHEV, Marandon, Marchegiani, Marcowith, Mares, Marti'i-Devesa,
  Marx, Maurin, Meintjes, Meyer, Mitchell, Moderski, Mohrmann, Montanari,
  Moore, Morris, Moulin, Muller, Murach, Nakashima, Naurois~(de), Nayerhoda,
  Ndiyavala, Niemiec, Noel, O'Brien, Oberholzer, Ohm, Olivera-Nieto,
  Ona-Wilhelmi~(de), Ostrowski, Panny, Panter, Parsons, Peron, Pita, Poireau,
  Prokhorov, Prokoph, PUEHLHOFER, Punch, Quirrenbach, Reichherzer, Reimer,
  Reimer, Remy, Renaud, Reville, Rieger, Romoli, Rowell, Rudak, Rueda~Ricarte,
  Ruiz~Velasco, Sahakian, Sailer, Salzmann, Sanchez, Santangelo, Sasaki,
  Schaefer, Schutte, Schwanke, Schüssler, Senniappan, Seyffert, Shapopi,
  Shiningayamwe, Simoni, Sinha, Sol, Spackman, Specovius, Spencer, Spir-Jacob,
  Stawarz, Steenkamp, Stegmann, Steinmassl, Steppa, Sun, Takahashi, Tanaka,
  Tavernier, Taylor, Terrier, Thiersen, Thorpe-Morgan, Tluczykont, Tomankova,
  Tsirou, Tsuji, Tuffs, Uchiyama, van~der Walt, van Eldik, van Rensburg, van
  Soelen, Vasileiadis, Veh, Venter, Vincent, Vink, Völk, Wagner, Watson,
  Werner, White, Wierzcholska, Wong, Yassin, Yusafzai, Zacharias, Zanin,
  Zargaryan, Zdziarski, Zech, Zhu, Zmija, Zouari, \& Zywucka}]{Acciari:2021YA}
Acciari, V.~A., Ansoldi, S., Antonelli, L.~A., {et~al.} 2021, in Proceedings of
  37th International Cosmic Ray Conference {\textemdash} PoS(ICRC2021), Vol.
  395, 960

\bibitem[{Aghanim {et~al.}(2020)}]{Aghanim:2018eyx}
Aghanim, N., {et~al.} 2020, Astron. Astrophys., 641, A6,
  \dodoi{10.1051/0004-6361/201833910}

\bibitem[{Ahlers \& Halzen(2018)}]{Ahlers:2018fkn}
Ahlers, M., \& Halzen, F. 2018, Prog. Part. Nucl. Phys., 102, 73,
  \dodoi{10.1016/j.ppnp.2018.05.001}

\bibitem[{Albert {et~al.}(2021)}]{ANTARES:2021jmp}
Albert, A., {et~al.} 2021.
\newblock \doarXiv{2103.15526}

\bibitem[{Anchordoqui {et~al.}(2014)}]{Anchordoqui:2013dnh}
Anchordoqui, L.~A., {et~al.} 2014, JHEAp, 1-2, 1,
  \dodoi{10.1016/j.jheap.2014.01.001}

\bibitem[{{Auchettl} {et~al.}(2017){Auchettl}, {Guillochon}, \&
  {Ramirez-Ruiz}}]{2017ApJ...838..149A}
{Auchettl}, K., {Guillochon}, J., \& {Ramirez-Ruiz}, E. 2017, \apj, 838, 149,
  \dodoi{10.3847/1538-4357/aa633b}

\bibitem[{Aydi {et~al.}(2020)}]{Aydi:2020znu}
Aydi, E., {et~al.} 2020, Nature Astron., 4, 776,
  \dodoi{10.1038/s41550-020-1070-y}

\bibitem[{{Bellm} {et~al.}(2019){Bellm}, {Kulkarni}, {Graham}, {Dekany},
  {Smith}, {Riddle}, {Masci}, {Helou}, {Prince}, {Adams}, {Barbarino},
  {Barlow}, {Bauer}, {Beck}, {Belicki}, {Biswas}, {Blagorodnova}, {Bodewits},
  {Bolin}, {Brinnel}, {Brooke}, {Bue}, {Bulla}, {Burruss}, {Cenko}, {Chang},
  {Connolly}, {Coughlin}, {Cromer}, {Cunningham}, {De}, {Delacroix}, {Desai},
  {Duev}, {Eadie}, {Farnham}, {Feeney}, {Feindt}, {Flynn}, {Franckowiak},
  {Frederick}, {Fremling}, {Gal-Yam}, {Gezari}, {Giomi}, {Goldstein},
  {Golkhou}, {Goobar}, {Groom}, {Hacopians}, {Hale}, {Henning}, {Ho}, {Hover},
  {Howell}, {Hung}, {Huppenkothen}, {Imel}, {Ip}, {Ivezi{\'c}}, {Jackson},
  {Jones}, {Juric}, {Kasliwal}, {Kaspi}, {Kaye}, {Kelley}, {Kowalski},
  {Kramer}, {Kupfer}, {Landry}, {Laher}, {Lee}, {Lin}, {Lin}, {Lunnan},
  {Giomi}, {Mahabal}, {Mao}, {Miller}, {Monkewitz}, {Murphy}, {Ngeow},
  {Nordin}, {Nugent}, {Ofek}, {Patterson}, {Penprase}, {Porter}, {Rauch},
  {Rebbapragada}, {Reiley}, {Rigault}, {Rodriguez}, {van Roestel}, {Rusholme},
  {van Santen}, {Schulze}, {Shupe}, {Singer}, {Soumagnac}, {Stein}, {Surace},
  {Sollerman}, {Szkody}, {Taddia}, {Terek}, {Van Sistine}, {van Velzen},
  {Vestrand}, {Walters}, {Ward}, {Ye}, {Yu}, {Yan}, \&
  {Zolkower}}]{2019PASP..131a8002B}
{Bellm}, E.~C., {Kulkarni}, S.~R., {Graham}, M.~J., {et~al.} 2019, Pub. Astron.
  Soc. Pacific, 131, 018002, \dodoi{10.1088/1538-3873/aaecbe}

\bibitem[{{Brown} {et~al.}(2017){Brown}, {Holoien}, {Auchettl}, {Stanek},
  {Kochanek}, {Shappee}, {Prieto}, \& {Grupe}}]{2017MNRAS.466.4904B}
{Brown}, J.~S., {Holoien}, T.~W.~S., {Auchettl}, K., {et~al.} 2017, \mnras,
  466, 4904, \dodoi{10.1093/mnras/stx033}

\bibitem[{Caprioli {et~al.}(2021)Caprioli, Haggerty, \&
  Blasi}]{Caprioli:2021nwv}
Caprioli, D., Haggerty, C., \& Blasi, P. 2021, PoS, ICRC2021, 482,
  \dodoi{10.22323/1.395.0482}

\bibitem[{Caprioli \& Spitkovsky(2014)}]{Caprioli:2013dca}
Caprioli, D., \& Spitkovsky, A. 2014, Astrophys. J., 783, 91,
  \dodoi{10.1088/0004-637X/783/2/91}

\bibitem[{Cardillo {et~al.}(2015)Cardillo, Amato, \& Blasi}]{Cardillo:2015zda}
Cardillo, M., Amato, E., \& Blasi, P. 2015, Astropart. Phys., 69, 1,
  \dodoi{10.1016/j.astropartphys.2015.03.002}

\bibitem[{Cendes {et~al.}(2021)Cendes, Alexander, Berger, Eftekhari, Williams,
  \& Chornock}]{2021arXiv210306299C}
Cendes, Y., Alexander, K.~D., Berger, E., {et~al.} 2021, Astrophys. J., 919,
  127, \dodoi{10.3847/1538-4357/ac110a}

\bibitem[{{Chandra}(2018)}]{Chandra2018}
{Chandra}, P. 2018, Sp. Sci. Rev., 214, 27, \dodoi{10.1007/s11214-017-0461-6}

\bibitem[{{Charalampopoulos} {et~al.}(2021){Charalampopoulos}, {Leloudas},
  {Malesani}, {Wevers}, {Arcavi}, {Nicholl}, {Pursiainen}, {Lawrence},
  {Anderson}, {Benetti}, {Cannizzaro}, {Chen}, {Galbany}, {Gromadzki},
  {Guti{\'e}rrez}, {Inserra}, {Jonker}, {M{\"u}ller-Bravo}, {Onori}, {Short},
  {Sollerman}, \& {Young}}]{Charalampopoulos2021}
{Charalampopoulos}, P., {Leloudas}, G., {Malesani}, D.~B., {et~al.} 2021, arXiv
  e-prints, arXiv:2109.00016.
\newblock \doarXiv{2109.00016}

\bibitem[{{Chatzopoulos} \& {Tuminello}(2019)}]{Chatzopoulos2019}
{Chatzopoulos}, E., \& {Tuminello}, R. 2019, Astrophys. J., 874, 68,
  \dodoi{10.3847/1538-4357/ab0ae6}

\bibitem[{{Chevalier}(1982)}]{1982ApJ...258..790C}
{Chevalier}, R.~A. 1982, \apj, 258, 790, \dodoi{10.1086/160126}

\bibitem[{{Chevalier} \& {Fransson}(1994)}]{1994ApJ...420..268C}
{Chevalier}, R.~A., \& {Fransson}, C. 1994, \apj, 420, 268,
  \dodoi{10.1086/173557}

\bibitem[{{Chevalier} \& {Fransson}(2003)}]{2003LNP...598..171C}
---. 2003, {Supernova Interaction with a Circumstellar Medium}, ed.
  K.~{Weiler}, Vol. 598, 171--194

\bibitem[{Chevalier \& Fransson(2016)}]{Chevalier:2016hzo}
Chevalier, R.~A., \& Fransson, C. 2016

\bibitem[{Chevalier \& Irwin(2011)}]{Chevalier:2011ha}
Chevalier, R.~A., \& Irwin, C.~M. 2011, Astrophys. J. Lett., 729, L6,
  \dodoi{10.1088/2041-8205/729/1/L6}

\bibitem[{{Chornock} {et~al.}(2019){Chornock}, {Blanchard}, {Gomez},
  {Hosseinzadeh}, \& {Berger}}]{Chornock2019}
{Chornock}, R., {Blanchard}, P.~K., {Gomez}, S., {Hosseinzadeh}, G., \&
  {Berger}, E. 2019, Transient Name Server Classification Report, 2019-1016, 1

\bibitem[{Dai \& Fang(2017)}]{Dai:2016gtz}
Dai, L., \& Fang, K. 2017, Mon. Not. Roy. Astron. Soc., 469, 1354,
  \dodoi{10.1093/mnras/stx863}

\bibitem[{Dessart {et~al.}(2015)Dessart, Audit, \& Hillier}]{Dessart:2015xaa}
Dessart, L., Audit, E., \& Hillier, D.~J. 2015, Mon. Not. Roy. Astron. Soc.,
  449, 4304, \dodoi{10.1093/mnras/stv609}

\bibitem[{{Ellison} {et~al.}(2007){Ellison}, {Patnaude}, {Slane}, {Blasi}, \&
  {Gabici}}]{2007ApJ...661..879E}
{Ellison}, D.~C., {Patnaude}, D.~J., {Slane}, P., {Blasi}, P., \& {Gabici}, S.
  2007, \apj, 661, 879, \dodoi{10.1086/517518}

\bibitem[{Esteban {et~al.}(2020)Esteban, Gonzalez-Garcia, Maltoni, Schwetz, \&
  Zhou}]{Esteban:2020cvm}
Esteban, I., Gonzalez-Garcia, M., Maltoni, M., Schwetz, T., \& Zhou, A. 2020,
  JHEP, 09, 178, \dodoi{10.1007/JHEP09(2020)178}

\bibitem[{Fang {et~al.}(2020)Fang, Metzger, Vurm, Aydi, \&
  Chomiuk}]{Fang:2020bkm}
Fang, K., Metzger, B.~D., Vurm, I., Aydi, E., \& Chomiuk, L. 2020, Astrophys.
  J., 904, 4, \dodoi{10.3847/1538-4357/abbc6e}

\bibitem[{{Finke} \& {Dermer}(2012)}]{2012ApJ...751...65F}
{Finke}, J.~D., \& {Dermer}, C.~D. 2012, Astrophys. J., 751, 65,
  \dodoi{10.1088/0004-637X/751/1/65}

\bibitem[{Franckowiak {et~al.}(2020)}]{Franckowiak:2020qrq}
Franckowiak, A., {et~al.} 2020, Astrophys. J., 893, 162,
  \dodoi{10.3847/1538-4357/ab8307}

\bibitem[{Franco {et~al.}(1992)Franco, Ferrini, \& Tenorio-Tagle}]{Franco_1992}
Franco, J., Ferrini, F., \& Tenorio-Tagle, G. 1992, in Proceedings of the 4th
  EIPC Astrophysical Workshop

\bibitem[{{Frederick} {et~al.}(2021){Frederick}, {Gezari}, {Graham},
  {Sollerman}, {van Velzen}, {Perley}, {Stern}, {Ward}, {Hammerstein}, {Hung},
  {Yan}, {Andreoni}, {Bellm}, {Duev}, {Kowalski}, {Mahabal}, {Masci},
  {Medford}, {Rusholme}, {Smith}, \& {Walters}}]{2020arXiv201008554F}
{Frederick}, S., {Gezari}, S., {Graham}, M.~J., {et~al.} 2021, \apj, 920, 56,
  \dodoi{10.3847/1538-4357/ac110f}

\bibitem[{Gal-Yam(2012)}]{Gal-Yam:2012ukv}
Gal-Yam, A. 2012, Science, 337, 927, \dodoi{10.1126/science.1203601}

\bibitem[{{Gal-Yam}(2017)}]{2017hsn..book..195G}
{Gal-Yam}, A. 2017, {Observational and Physical Classification of Supernovae},
  ed. A.~W. {Alsabti} \& P.~{Murdin}, 195

\bibitem[{{Gal-Yam}(2019)}]{Gal-Yam2019}
---. 2019, Ann. Rev. Astron. Astrophys., 57, 305,
  \dodoi{10.1146/annurev-astro-081817-051819}

\bibitem[{{Gall} {et~al.}(2014){Gall}, {Hjorth}, {Watson}, {Dwek}, {Maund},
  {Fox}, {Leloudas}, {Malesani}, \& {Day-Jones}}]{Gall2014}
{Gall}, C., {Hjorth}, J., {Watson}, D., {et~al.} 2014, \nat, 511, 326,
  \dodoi{10.1038/nature13558}

\bibitem[{Garrappa {et~al.}(2019)}]{Fermi-LAT:2019hte}
Garrappa, S., {et~al.} 2019, Astrophys. J., 880, 880:103,
  \dodoi{10.3847/1538-4357/ab2ada}

\bibitem[{{Gehrels} {et~al.}(2004){Gehrels}, {Chincarini}, {Giommi}, {Mason},
  {Nousek}, {Wells}, {White}, {Barthelmy}, {Burrows}, {Cominsky}, {Hurley},
  {Marshall}, {M{\'e}sz{\'a}ros}, {Roming}, {Angelini}, {Barbier}, {Belloni},
  {Campana}, {Caraveo}, {Chester}, {Citterio}, {Cline}, {Cropper}, {Cummings},
  {Dean}, {Feigelson}, {Fenimore}, {Frail}, {Fruchter}, {Garmire}, {Gendreau},
  {Ghisellini}, {Greiner}, {Hill}, {Hunsberger}, {Krimm}, {Kulkarni}, {Kumar},
  {Lebrun}, {Lloyd-Ronning}, {Markwardt}, {Mattson}, {Mushotzky}, {Norris},
  {Osborne}, {Paczynski}, {Palmer}, {Park}, {Parsons}, {Paul}, {Rees},
  {Reynolds}, {Rhoads}, {Sasseen}, {Schaefer}, {Short}, {Smale}, {Smith},
  {Stella}, {Tagliaferri}, {Takahashi}, {Tashiro}, {Townsley}, {Tueller},
  {Turner}, {Vietri}, {Voges}, {Ward}, {Willingale}, {Zerbi}, \&
  {Zhang}}]{gehrels04}
{Gehrels}, N., {Chincarini}, G., {Giommi}, P., {et~al.} 2004, \apj, 611, 1005,
  \dodoi{10.1086/422091}

\bibitem[{Giommi {et~al.}(2020)Giommi, Padovani, Oikonomou, Glauch, Paiano, \&
  Resconi}]{Giommi:2020viy}
Giommi, P., Padovani, P., Oikonomou, F., {et~al.} 2020, Astron. Astrophys.,
  640, L4, \dodoi{10.1051/0004-6361/202038423}

\bibitem[{{Hinkle} {et~al.}(2021){Hinkle}, {Holoien}, {Auchettl}, {Shappee},
  {Neustadt}, {Payne}, {Brown}, {Kochanek}, {Stanek}, {Graham}, {Tucker}, {Do},
  {Anderson}, {Bose}, {Chen}, {Coulter}, {Dimitriadis}, {Dong}, {Foley},
  {Huber}, {Hung}, {Kilpatrick}, {Pignata}, {Piro}, {Rojas-Bravo}, {Siebert},
  {Stalder}, {Thompson}, {Tonry}, {Vallely}, \&
  {Wisniewski}}]{2021MNRAS.500.1673H}
{Hinkle}, J.~T., {Holoien}, T.~W.~S., {Auchettl}, K., {et~al.} 2021, \mnras,
  500, 1673, \dodoi{10.1093/mnras/staa3170}

\bibitem[{{Holoien} {et~al.}(2019){Holoien}, {Vallely}, {Auchettl}, {Stanek},
  {Kochanek}, {French}, {Prieto}, {Shappee}, {Brown}, {Fausnaugh}, {Dong},
  {Thompson}, {Bose}, {Neustadt}, {Cacella}, {Brimacombe}, {Kendurkar},
  {Beaton}, {Boutsia}, {Chomiuk}, {Connor}, {Morrell}, {Newman}, {Rudie},
  {Shishkovksy}, \& {Strader}}]{2019ApJ...883..111H}
{Holoien}, T. W.~S., {Vallely}, P.~J., {Auchettl}, K., {et~al.} 2019, \apj,
  883, 111, \dodoi{10.3847/1538-4357/ab3c66}

\bibitem[{{Hung} {et~al.}(2017){Hung}, {Gezari}, {Blagorodnova}, {Roth},
  {Cenko}, {Kulkarni}, {Horesh}, {Arcavi}, {McCully}, {Yan}, {Lunnan},
  {Fremling}, {Cao}, {Nugent}, \& {Wozniak}}]{2017ApJ...842...29H}
{Hung}, T., {Gezari}, S., {Blagorodnova}, N., {et~al.} 2017, \apj, 842, 29,
  \dodoi{10.3847/1538-4357/aa7337}

\bibitem[{{Hung} {et~al.}(2020){Hung}, {Foley}, {Ramirez-Ruiz}, {Dai},
  {Auchettl}, {Kilpatrick}, {Mockler}, {Brown}, {Coulter}, {Dimitriadis},
  {Holoien}, {Law-Smith}, {Piro}, {Rest}, {Rojas-Bravo}, \&
  {Siebert}}]{2020ApJ...903...31H}
{Hung}, T., {Foley}, R.~J., {Ramirez-Ruiz}, E., {et~al.} 2020, \apj, 903, 31,
  \dodoi{10.3847/1538-4357/abb606}

\bibitem[{{Hung} {et~al.}(2021){Hung}, {Foley}, {Veilleux}, {Cenko}, {Dai},
  {Auchettl}, {Brink}, {Dimitriadis}, {Filippenko}, {Gezari}, {Holoien},
  {Kilpatrick}, {Mockler}, {Piro}, {Ramirez-Ruiz}, {Rojas-Bravo}, {Siebert},
  {van Velzen}, \& {Zheng}}]{2021ApJ...917....9H}
{Hung}, T., {Foley}, R.~J., {Veilleux}, S., {et~al.} 2021, \apj, 917, 9,
  \dodoi{10.3847/1538-4357/abf4c3}

\bibitem[{{IceCube Collaboration}(2020)}]{2020GCN.27865....1I}
{IceCube Collaboration}. 2020, GRB Coordinates Network, 27865, 1

\bibitem[{{Jonker} {et~al.}(2020){Jonker}, {Stone}, {Generozov}, {van Velzen},
  \& {Metzger}}]{2020ApJ...889..166J}
{Jonker}, P.~G., {Stone}, N.~C., {Generozov}, A., {van Velzen}, S., \&
  {Metzger}, B. 2020, \apj, 889, 166, \dodoi{10.3847/1538-4357/ab659c}

\bibitem[{Kadler {et~al.}(2016)}]{Kadler:2016ygj}
Kadler, M., {et~al.} 2016, Nature Phys., 12, 807, \dodoi{10.1038/NPHYS3715}

\bibitem[{Katz {et~al.}(2011)Katz, Sapir, \& Waxman}]{Katz:2011zx}
Katz, B., Sapir, N., \& Waxman, E. 2011.
\newblock \doarXiv{1106.1898}

\bibitem[{Kelner {et~al.}(2006)Kelner, Aharonian, \& Bugayov}]{Kelner:2006tc}
Kelner, S.~R., Aharonian, F.~A., \& Bugayov, V.~V. 2006, Phys. Rev. D, 74,
  034018, \dodoi{10.1103/PhysRevD.74.034018}

\bibitem[{Krau\ss{} {et~al.}(2018)Krau\ss{}, Deoskar, Baxter, Kadler, Kreter,
  Langejahn, Mannheim, Polko, Wang, \& Wilms}]{Krauss:2018tpa}
Krau\ss{}, F., Deoskar, K., Baxter, C., {et~al.} 2018, Astron. Astrophys., 620,
  A174, \dodoi{10.1051/0004-6361/201834183}

\bibitem[{Levinson \& Bromberg(2008)}]{Levinson:2007rj}
Levinson, A., \& Bromberg, O. 2008, Phys. Rev. Lett., 100, 131101,
  \dodoi{10.1103/PhysRevLett.100.131101}

\bibitem[{Liu {et~al.}(2020)Liu, Xi, \& Wang}]{Liu:2020isi}
Liu, R.-Y., Xi, S.-Q., \& Wang, X.-Y. 2020, Phys. Rev. D, 102, 083028,
  \dodoi{10.1103/PhysRevD.102.083028}

\bibitem[{Lunardini \& Winter(2017)}]{Lunardini:2016xwi}
Lunardini, C., \& Winter, W. 2017, Phys. Rev. D, 95, 123001,
  \dodoi{10.1103/PhysRevD.95.123001}

\bibitem[{{Matsumoto} \& {Piran}(2021)}]{2021MNRAS.507.4196M}
{Matsumoto}, T., \& {Piran}, T. 2021, Mon. Not. Roy. Astron. Soc., 507, 4196,
  \dodoi{10.1093/mnras/stab2418}

\bibitem[{{Matsumoto} {et~al.}(2021){Matsumoto}, {Piran}, \&
  {Krolik}}]{2021arXiv210902648M}
{Matsumoto}, T., {Piran}, T., \& {Krolik}, J.~H. 2021, arXiv e-prints,
  arXiv:2109.02648.
\newblock \doarXiv{2109.02648}

\bibitem[{M\'esz\'aros(2017{\natexlab{a}})}]{Meszaros:2017fcs}
M\'esz\'aros, P. 2017{\natexlab{a}}, Ann. Rev. Nucl. Part. Sci., 67, 45,
  \dodoi{10.1146/annurev-nucl-101916-123304}

\bibitem[{M\'esz\'aros(2017{\natexlab{b}})}]{Meszaros:2015krr}
---. 2017{\natexlab{b}}, {Gamma Ray Bursts as Neutrino Sources}

\bibitem[{{Mohan} {et~al.}(2021){Mohan}, {An}, {Zhang}, {Yang}, {Yang}, \&
  {Wang}}]{2021arXiv210615799M}
{Mohan}, P., {An}, T., {Zhang}, Y., {et~al.} 2021, arXiv e-prints,
  arXiv:2106.15799.
\newblock \doarXiv{2106.15799}

\bibitem[{Moriya {et~al.}(2013)Moriya, Maeda, Taddia, Sollerman, Blinnikov, \&
  Sorokina}]{Moriya:2013hka}
Moriya, T.~J., Maeda, K., Taddia, F., {et~al.} 2013, Mon. Not. Roy. Astron.
  Soc., 435, 1520, \dodoi{10.1093/mnras/stt1392}

\bibitem[{Moriya {et~al.}(2014)Moriya, Maeda, Taddia, Sollerman, Blinnikov, \&
  Sorokina}]{Moriya:2014cua}
---. 2014, Mon. Not. Roy. Astron. Soc., 439, 2917, \dodoi{10.1093/mnras/stu163}

\bibitem[{Moriya {et~al.}(2018)Moriya, Sorokina, \& Chevalier}]{Moriya:2018sig}
Moriya, T.~J., Sorokina, E.~I., \& Chevalier, R.~A. 2018, Space Sci. Rev., 214,
  59, \dodoi{10.1007/s11214-018-0493-6}

\bibitem[{Murase(2017)}]{Murase:2015ndr}
Murase, K. 2017, {Active Galactic Nuclei as High-Energy Neutrino Sources}, ed.
  T.~Gaisser \& A.~Karle

\bibitem[{Murase(2018)}]{Murase:2017pfe}
---. 2018, Phys. Rev. D, 97, 081301, \dodoi{10.1103/PhysRevD.97.081301}

\bibitem[{Murase {et~al.}(2020)Murase, Kimura, Zhang, Oikonomou, \&
  Petropoulou}]{Murase:2020lnu}
Murase, K., Kimura, S.~S., Zhang, B.~T., Oikonomou, F., \& Petropoulou, M.
  2020, Astrophys. J., 902, 108, \dodoi{10.3847/1538-4357/abb3c0}

\bibitem[{Murase {et~al.}(2011)Murase, Thompson, Lacki, \&
  Beacom}]{Murase:2010cu}
Murase, K., Thompson, T.~A., Lacki, B.~C., \& Beacom, J.~F. 2011, Phys. Rev. D,
  84, 043003, \dodoi{10.1103/PhysRevD.84.043003}

\bibitem[{{Murase} {et~al.}(2014){Murase}, {Thompson}, \&
  {Ofek}}]{2014MNRAS.440.2528M}
{Murase}, K., {Thompson}, T.~A., \& {Ofek}, E.~O. 2014, Mon. Not. Roy. Astron.
  Soc., 440, 2528, \dodoi{10.1093/mnras/stu384}

\bibitem[{{Nyholm} {et~al.}(2017){Nyholm}, {Sollerman}, {Taddia}, {Fremling},
  {Moriya}, {Ofek}, {Gal-Yam}, {De Cia}, {Roy}, {Kasliwal}, {Cao}, {Nugent}, \&
  {Masci}}]{Nyholm2017}
{Nyholm}, A., {Sollerman}, J., {Taddia}, F., {et~al.} 2017, Astron. Astrophys.,
  605, A6, \dodoi{10.1051/0004-6361/201629906}

\bibitem[{{Nyholm} {et~al.}(2020){Nyholm}, {Sollerman}, {Tartaglia}, {Taddia},
  {Fremling}, {Blagorodnova}, {Filippenko}, {Gal-Yam}, {Howell},
  {Karamehmetoglu}, {Kulkarni}, {Laher}, {Leloudas}, {Masci}, {Kasliwal},
  {Mor{\r{a}}}, {Moriya}, {Ofek}, {Papadogiannakis}, {Quimby}, {Rebbapragada},
  \& {Schulze}}]{Nyholm2020}
{Nyholm}, A., {Sollerman}, J., {Tartaglia}, L., {et~al.} 2020, Astron.
  Astrophys., 637, A73, \dodoi{10.1051/0004-6361/201936097}

\bibitem[{Pan {et~al.}(2013)Pan, Patnaude, \& Loeb}]{Pan:2013nfa}
Pan, T., Patnaude, D.~J., \& Loeb, A. 2013, Mon. Not. Roy. Astron. Soc., 433,
  838, \dodoi{10.1093/mnras/stt780}

\bibitem[{Patnaude \& Fesen(2009)}]{Patnaude:2008gq}
Patnaude, D.~J., \& Fesen, R.~A. 2009, Astrophys. J., 697, 535,
  \dodoi{10.1088/0004-637X/697/1/535}

\bibitem[{{Patterson} {et~al.}(2019){Patterson}, {Bellm}, {Rusholme}, {Masci},
  {Juric}, {Krughoff}, {Golkhou}, {Graham}, {Kulkarni}, {Helou}, \& {Zwicky
  Transient Facility Collaboration}}]{Patterson2019}
{Patterson}, M.~T., {Bellm}, E.~C., {Rusholme}, B., {et~al.} 2019, Publ.
  Astron. Soc. Pac., 131, 018001, \dodoi{10.1088/1538-3873/aae904}

\bibitem[{Petropoulou {et~al.}(2017)Petropoulou, Coenders, Vasilopoulos,
  Kamble, \& Sironi}]{Petropoulou:2017ymv}
Petropoulou, M., Coenders, S., Vasilopoulos, G., Kamble, A., \& Sironi, L.
  2017, Mon. Not. Roy. Astron. Soc., 470, 1881, \dodoi{10.1093/mnras/stx1251}

\bibitem[{Petropoulou {et~al.}(2016)Petropoulou, Kamble, \&
  Sironi}]{Petropoulou:2016zar}
Petropoulou, M., Kamble, A., \& Sironi, L. 2016, Mon. Not. Roy. Astron. Soc.,
  460, 44, \dodoi{10.1093/mnras/stw920}

\bibitem[{Pitik {et~al.}(2021)Pitik, Tamborra, \& Petropoulou}]{Pitik:2021xhb}
Pitik, T., Tamborra, I., \& Petropoulou, M. 2021, JCAP, 05, 034,
  \dodoi{10.1088/1475-7516/2021/05/034}

\bibitem[{Protheroe \& Clay(2004)}]{Protheroe:2003vc}
Protheroe, R.~J., \& Clay, R.~W. 2004, Publ. Astron. Soc. Pac., 21, 1,
  \dodoi{10.1071/AS03047}

\bibitem[{Quimby {et~al.}(2013)Quimby, Yuan, Akerlof, \&
  Wheeler}]{Quimby:2013jb}
Quimby, R.~M., Yuan, F., Akerlof, C., \& Wheeler, J.~C. 2013, Mon. Not. Roy.
  Astron. Soc., 431, 912, \dodoi{10.1093/mnras/stt213}

\bibitem[{{Rees}(1988)}]{Rees1988}
{Rees}, M.~J. 1988, Nature, 333, 523, \dodoi{10.1038/333523a0}

\bibitem[{{Reusch} {et~al.}(2020{\natexlab{a}}){Reusch}, {Stein},
  {Franckowiak}, {Gezari}, {Zwicky Transient Facility (Ztf) Collaboration}, \&
  {Global Relay Of Observatories Watching Transients Happen (Growth)
  Collaboration}}]{2020GCN.27872....1R}
{Reusch}, S., {Stein}, R., {Franckowiak}, A., {et~al.} 2020{\natexlab{a}}, GRB
  Coordinates Network, 27872, 1

\bibitem[{{Reusch} {et~al.}(2020{\natexlab{b}}){Reusch}, {Stein},
  {Franckowiak}, {Necker}, {Sollerman}, {Barbarino}, \&
  {Schweyer}}]{2020GCN.27980....1R}
---. 2020{\natexlab{b}}, GRB Coordinates Network, 27980, 1

\bibitem[{{Reusch} {et~al.}(2020{\natexlab{c}}){Reusch}, {Stein},
  {Franckowiak}, {Sollerman}, {Schweyer}, \& {Barbarino}}]{2020GCN.27910....1R}
---. 2020{\natexlab{c}}, GRB Coordinates Network, 27910, 1

\bibitem[{Reusch {et~al.}(2021)}]{Reusch:2021ztx}
Reusch, S., {et~al.} 2021.
\newblock \doarXiv{2111.09390}

\bibitem[{{Ricci} {et~al.}(2017){Ricci}, {Trakhtenbrot}, {Koss}, {Ueda}, {Del
  Vecchio}, {Treister}, {Schawinski}, {Paltani}, {Oh}, {Lamperti}, {Berney},
  {Gandhi}, {Ichikawa}, {Bauer}, {Ho}, {Asmus}, {Beckmann}, {Soldi},
  {Balokovi{\'c}}, {Gehrels}, \& {Markwardt}}]{2017ApJS..233...17R}
{Ricci}, C., {Trakhtenbrot}, B., {Koss}, M.~J., {et~al.} 2017, \apjs, 233, 17,
  \dodoi{10.3847/1538-4365/aa96ad}

\bibitem[{{Sato} {et~al.}(2018){Sato}, {Katsuda}, {Morii}, {Bamba}, {Hughes},
  {Maeda}, {Ishida}, \& {Fraschetti}}]{2018ApJ...853...46S}
{Sato}, T., {Katsuda}, S., {Morii}, M., {et~al.} 2018, \apj, 853, 46,
  \dodoi{10.3847/1538-4357/aaa021}

\bibitem[{{Schure} {et~al.}(2010){Schure}, {Achterberg}, {Keppens}, \&
  {Vink}}]{2010MNRAS.406.2633S}
{Schure}, K.~M., {Achterberg}, A., {Keppens}, R., \& {Vink}, J. 2010, \mnras,
  406, 2633, \dodoi{10.1111/j.1365-2966.2010.16857.x}

\bibitem[{Senno {et~al.}(2017)Senno, Murase, \& M\'esz\'aros}]{Senno:2016bso}
Senno, N., Murase, K., \& M\'esz\'aros, P. 2017, Astrophys. J., 838, 3,
  \dodoi{10.3847/1538-4357/aa6344}

\bibitem[{{Slane} {et~al.}(2015){Slane}, {Lee}, {Ellison}, {Patnaude},
  {Hughes}, {Eriksen}, {Castro}, \& {Nagataki}}]{2015ApJ...799..238S}
{Slane}, P., {Lee}, S.~H., {Ellison}, D.~C., {et~al.} 2015, \apj, 799, 238,
  \dodoi{10.1088/0004-637X/799/2/238}

\bibitem[{{Smith} {et~al.}(2020){Smith}, {Smartt}, {Young}, {Tonry}, {Denneau},
  {Flewelling}, {Heinze}, {Weiland}, {Stalder}, {Rest}, {Stubbs}, {Anderson},
  {Chen}, {Clark}, {Do}, {F{\"o}rster}, {Fulton}, {Gillanders}, {McBrien},
  {O'Neill}, {Srivastav}, \& {Wright}}]{ATLAS2020}
{Smith}, K.~W., {Smartt}, S.~J., {Young}, D.~R., {et~al.} 2020, Publ. Astron.
  Soc. Pac., 132, 085002, \dodoi{10.1088/1538-3873/ab936e}

\bibitem[{Smith(2014)}]{Smith:2014txa}
Smith, N. 2014, Ann. Rev. Astron. Astrophys., 52, 487,
  \dodoi{10.1146/annurev-astro-081913-040025}

\bibitem[{Stein(2021)}]{Stein:2021FH}
Stein, R. 2021, in Proceedings of 37th International Cosmic Ray Conference
  {\textemdash} PoS(ICRC2021), Vol. 395, 009

\bibitem[{Stein {et~al.}(2021)}]{Stein:2020xhk}
Stein, R., {et~al.} 2021, Nature Astron., 5, 510,
  \dodoi{10.1038/s41550-020-01295-8}

\bibitem[{{Stritzinger} {et~al.}(2012){Stritzinger}, {Taddia}, {Fransson},
  {Fox}, {Morrell}, {Phillips}, {Sollerman}, {Anderson}, {Boldt}, {Brown},
  {Campillay}, {Castellon}, {Contreras}, {Folatelli}, {Habergham}, {Hamuy},
  {Hjorth}, {James}, {Krzeminski}, {Mattila}, {Persson}, \&
  {Roth}}]{Stritzinger2012}
{Stritzinger}, M., {Taddia}, F., {Fransson}, C., {et~al.} 2012, Astrophys. J.,
  756, 173, \dodoi{10.1088/0004-637X/756/2/173}

\bibitem[{Strotjohann {et~al.}(2019)Strotjohann, Kowalski, \&
  Franckowiak}]{Strotjohann:2018ufz}
Strotjohann, N.~L., Kowalski, M., \& Franckowiak, A. 2019, Astron. Astrophys.,
  622, L9, \dodoi{10.1051/0004-6361/201834750}

\bibitem[{{Sturner} {et~al.}(1997){Sturner}, {Skibo}, {Dermer}, \&
  {Mattox}}]{1997ApJ...490..619S}
{Sturner}, S.~J., {Skibo}, J.~G., {Dermer}, C.~D., \& {Mattox}, J.~R. 1997,
  Astrophys. J., 490, 619, \dodoi{10.1086/304894}

\bibitem[{Suvorova {et~al.}(2021)Suvorova, Allakhverdyan, Avrorin, Avrorin,
  Aynutdinov, Bannasch, Bard??ová, Belolaptikov, Borina, Brudanin, Budnev,
  Dik, Domogatsky, Doroshenko, Dvornický, Dyachok, Dzhilkibaev, Eckerová,
  Elzhov, Fajt, Fialkovski, Gafarov, Golubkov, Gorshkov, Gress, Katulin,
  Kebkal, Kebkal, Khramov, Kolbin, Konischev, Kopa?ski, Korobchenko,
  Koshechkin, Kozhin, Kruglov, Kryukov, Kulepov, Malecki, Malyshkin, Milenin,
  Mirgazov, Naumov, Nazari, Noga, Petukhov, Pliskovsky, Rozanov, Rushay,
  Ryabov, Safronov, Shaybonov, Shelepov, ¦imkovic, Sirenko, Skurikhin,
  Solovjev, Sorokovikov, ¦tekl, Stromakov, Sushenok, Tabolenko, Tarashansky,
  Yablokova, Yakovlev, \& Zaborov}]{Suvorova:2021ou}
Suvorova, O., Allakhverdyan, V., Avrorin, A., {et~al.} 2021, in Proceedings of
  37th International Cosmic Ray Conference {\textemdash} PoS(ICRC2021), Vol.
  395, 946

\bibitem[{{Suzuki} {et~al.}(2019){Suzuki}, {Moriya}, \&
  {Takiwaki}}]{Suzuki:2019kny}
{Suzuki}, A., {Moriya}, T.~J., \& {Takiwaki}, T. 2019, Astrophys. J., 887, 249,
  \dodoi{10.3847/1538-4357/ab5a83}

\bibitem[{Suzuki {et~al.}(2020)Suzuki, Moriya, \& Takiwaki}]{Suzuki:2020qui}
Suzuki, A., Moriya, T.~J., \& Takiwaki, T. 2020, Astrophys. J., 899, 56,
  \dodoi{10.3847/1538-4357/aba0ba}

\bibitem[{Suzuki {et~al.}(2021)Suzuki, Nicholl, Moriya, \&
  Takiwaki}]{Suzuki:2020yhz}
Suzuki, A., Nicholl, M., Moriya, T.~J., \& Takiwaki, T. 2021, Astrophys. J.,
  908, 99, \dodoi{10.3847/1538-4357/abd6ce}

\bibitem[{{Taddia} {et~al.}(2013){Taddia}, {Stritzinger}, {Sollerman},
  {Phillips}, {Anderson}, {Boldt}, {Campillay}, {Castell{\'o}n}, {Contreras},
  {Folatelli}, {Hamuy}, {Heinrich-Josties}, {Krzeminski}, {Morrell}, {Burns},
  {Freedman}, {Madore}, {Persson}, \& {Suntzeff}}]{Taddia2013}
{Taddia}, F., {Stritzinger}, M.~D., {Sollerman}, J., {et~al.} 2013, Astron.
  Astrophys., 555, A10, \dodoi{10.1051/0004-6361/201321180}

\bibitem[{Tamborra {et~al.}(2014)Tamborra, Ando, \& Murase}]{Tamborra:2014xia}
Tamborra, I., Ando, S., \& Murase, K. 2014, JCAP, 09, 043,
  \dodoi{10.1088/1475-7516/2014/09/043}

\bibitem[{{Tonry} {et~al.}(2018){Tonry}, {Denneau}, {Heinze}, {Stalder},
  {Smith}, {Smartt}, {Stubbs}, {Weiland}, \& {Rest}}]{ATLAS2018}
{Tonry}, J.~L., {Denneau}, L., {Heinze}, A.~N., {et~al.} 2018, Publ. Astron.
  Soc. Pac., 130, 064505, \dodoi{10.1088/1538-3873/aabadf}

\bibitem[{Vitagliano {et~al.}(2020)Vitagliano, Tamborra, \&
  Raffelt}]{Vitagliano:2019yzm}
Vitagliano, E., Tamborra, I., \& Raffelt, G. 2020, Rev. Mod. Phys., 92, 45006,
  \dodoi{10.1103/RevModPhys.92.045006}

\bibitem[{Wang \& Liu(2016)}]{Wang:2015mmh}
Wang, X.-Y., \& Liu, R.-Y. 2016, Phys. Rev. D, 93, 083005,
  \dodoi{10.1103/PhysRevD.93.083005}

\bibitem[{Waxman(2017)}]{Waxman:2015ues}
Waxman, E. 2017, {The Origin of IceCube\textquoteright{}s Neutrinos: Cosmic Ray
  Accelerators Embedded in Star Forming Calorimeters}

\bibitem[{{Weaver}(1976)}]{1976ApJS...32..233W}
{Weaver}, T.~A. 1976, \apjs, 32, 233, \dodoi{10.1086/190398}

\bibitem[{{Wevers} {et~al.}(2021){Wevers}, {Pasham}, {van Velzen},
  {Miller-Jones}, {Uttley}, {Gendreau}, {Remillard}, {Arzoumanian},
  {L{\"o}wenstein}, \& {Chiti}}]{2021ApJ...912..151W}
{Wevers}, T., {Pasham}, D.~R., {van Velzen}, S., {et~al.} 2021, \apj, 912, 151,
  \dodoi{10.3847/1538-4357/abf5e2}

\bibitem[{Winter \& Lunardini(2021)}]{Winter:2020ptf}
Winter, W., \& Lunardini, C. 2021, Nature Astron., 5, 5,
  \dodoi{10.1038/s41550-021-01343-x}

\bibitem[{{Yan} {et~al.}(2019){Yan}, {Perley}, {Lunnan}, {Schulze}, {Gal-Yam},
  {Taggart}, {Yaron}, \& {Velzen}}]{Yan:2019TNSAN}
{Yan}, L., {Perley}, D., {Lunnan}, R., {et~al.} 2019, Transient Name Server
  AstroNote, 45, 1

\bibitem[{Yuksel {et~al.}(2008)Yuksel, Kistler, Beacom, \&
  Hopkins}]{Yuksel:2008cu}
Yuksel, H., Kistler, M.~D., Beacom, J.~F., \& Hopkins, A.~M. 2008, Astrophys.
  J. Lett., 683, L5, \dodoi{10.1086/591449}

\bibitem[{Zandanel {et~al.}(2015)Zandanel, Tamborra, Gabici, \&
  Ando}]{Zandanel:2014pva}
Zandanel, F., Tamborra, I., Gabici, S., \& Ando, S. 2015, Astron. Astrophys.,
  578, A32, \dodoi{10.1051/0004-6361/201425249}

\bibitem[{Zirakashvili \& Ptuskin(2016)}]{Zirakashvili:2015mua}
Zirakashvili, V.~N., \& Ptuskin, V.~S. 2016, Astropart. Phys., 78, 28,
  \dodoi{10.1016/j.astropartphys.2016.02.004}

\bibitem[{Zyla {et~al.}(2020)}]{Zyla:2020zbs}
Zyla, P., {et~al.} 2020, PTEP, 2020, 083C01, \dodoi{10.1093/ptep/ptaa104}

\end{thebibliography}
\bibliographystyle{aasjournal}

\end{document}